\documentclass[useAMS,usenatbib]{mn2e}
\usepackage{natbib}
\usepackage{epsfig}
\usepackage{amsbsy}
\usepackage{hyperref}
\providecommand{\adsurl}[1]{\href{#1}{ADS}}

\newcommand{\lya}{Lyman-$\alpha$~}

\newcommand{\be}{\begin{equation}}
\newcommand{\ee}{\end{equation}}
\newcommand{\ba}{\begin{eqnarray}}
\newcommand{\ea}{\end{eqnarray}}
\newcommand{\brr}{\begin{array}}

\newcommand{\err}{\end{array}}
\newcommand{\bc}{\begin{center}}
\newcommand{\ec}{\end{center}}

\newcommand{\lb}{{\left<\right.}}
\newcommand{\rb}{{\left.\right>}}
\newcommand{\hm}{\,h^{-1}{\rm Mpc}}
\newcommand{\vel}{\,{\rm km\,s^{-1}}}

\newcommand{\msun}{\,h^{-1}M_\odot}
\newcommand{\mincir}{\raise
  -2.truept\hbox{\rlap{\hbox{$\sim$}}\raise5.truept \hbox{$<$}\ }}
\newcommand{\magcir}{\raise
  -2.truept\hbox{\rlap{\hbox{$\sim$}}\raise5.truept \hbox{$>$}\ }}

\newcommand{\db}{\mbox{$\delta_{\rm b}$}}

\DeclareMathAlphabet{\mathsc}{OT1}{cmr}{m}{sc}
\def\testbx{bx}%
\DeclareRobustCommand{\ion}[2]{%
\relax\ifmmode
\ifx\testbx\f@series
{\mathbf{#1\,\mathsc{#2}}}\else
{\mathrm{#1\,\mathsc{#2}}}\fi
\else\textup{#1\,{\mdseries\textsc{#2}}}%
\fi}
\title[The impact of feedback on the  low redshift IGM]
{The impact of feedback  on the low redshift Intergalactic Medium}

\author[Tornatore, Borgani, Viel \& Springel] 
{L. Tornatore$^{1,2,3}$, S.  Borgani$^{1,2,3}$,
M. Viel$^{2,3}$ \& V. Springel $^{4}$\\
$^1$ Dipartimento di Astronomia dell'Universit\`a di Trieste,  Via G.B. Tiepolo 11,
I-34131 Trieste, Italy\\
$^2$ INAF - Osservatorio Astronomico di Trieste, Via G.B. Tiepolo 11,
I-34131 Trieste, Italy \\
$^3$ INFN/National Institute for Nuclear Physics, Via Valerio 2,
I-34127 Trieste, Italy\\
$^4$  Max-Planck Institut fuer Astrophysik,
Karl-Schwarzschild Strasse 1, D-85748 Garching, Germany
\\}

\begin{document}
\maketitle
\begin{abstract}
  We analyse the evolution of the properties of the low-redshift
  Intergalactic Medium (IGM) using high-resolution hydrodynamic
  simulations that include a detailed chemical evolution model. We
  focus on the effects that two different forms of energy feedback,
  strong galactic winds driven by supernova explosion and Active
  Galactic Nuclei (AGN) powered by gas accretion onto super-massive
  black holes (BHs), have on the thermo- and chemo-dynamical
  properties of of the low redshift IGM. We find that feedback
  associated to winds {(W)} and BHs leave distinct signatures in
  both the chemical and thermal history of the baryons, especially at
  redshift $z<3$. { BH feedback produces an amount of gas with
    temperature in the range $10^5-10^7$K, the Warm Hot Intergalactic
    Medium (WHIM), larger than that produced by the wind feedback. At
    $z=0$ the fraction of baryons in the WHIM is about 50 per
    cent in the runs with BH feedback and about 40 per cent in the
    runs with wind feedback. 
The amount of warm baryons ($10^4<T<10^5$K) is instead at
    about the same level, $\sim 30$ per cent, in the runs with BH and
    wind feedback. Also, BH feedback provides a stronger and more
    pristine enrichment of the WHIM.}  We find that the metal--mass
  weighted age of WHIM enrichment at $z=0$ is on average a factor
  $\sim 1.5$ smaller in the BH run than for the corresponding runs
  with galactic winds. We present results for the enrichment in terms
  of mass and metallicity distributions for the WHIM phase, both as a
  function of density and temperature.  Finally, we compute the
  evolution of the relative abundances between different heavy
  elements, namely Oxygen, Carbon and Iron. While both C/O and O/Fe
  evolve differently at high redshifts for different feedback models,
  their values are similar at $z=0$. We also find that changing the
  stellar initial mass function has a smaller effect on the evolution
  of the above relative abundances than changing the feedback
  model. The sensitivity of WHIM properties on the implemented
  feedback scheme could be important both for discriminating between
  different feedback physics and for detecting the WHIM with future
  far-UV and X-ray telescopes.
\end{abstract}


\section{Introduction}
Solving the problem of the missing baryons at low redshift is one of
the important goals of observational cosmology
\citep[e.g.,][]{persicsalucci92,fukugita98,fukugita04}. Since the
pioneering work by \cite{cenostriker99}, cosmological hydrodynamic
simulations provided a fundamental contribution in this field
\citep[see also][]{dave01}. Although the baryon budget is closed by
observations at high redshift, $z\magcir 2$, simulations indicate that
about half of the baryons at low-$z$ should lie in a tenuous quite
elusive phase, the so-called Warm-Hot Integalactic Medium (WHIM). The
WHIM is predicted to be made of low density ($n_{\rm H}\sim
10^{-6}-10^{-4}$cm$^{-3}$) and relatively high temperature ($T\sim
10^5-10^7$K) plasma, whose main constituents are ionised Hydrogen and
Helium, with traces of heavier elements in high ionization
states. Simulations have also shown that gas at these densities traces
the filamentary structures which define the skeleton of the
large-scale cosmic web, and is heated by shocks from supersonic
gravitational accretion onto the forming potential wells.

Due to its low density, the collisionally ionised WHIM should be
characterised in emission by a very low surface brightness in the UV
and soft X--ray bands, so that its detection lies beyond the
capability of available detectors and should await for instrumentation
of the next generation \citep[e.g.,][]{yoshikawa04}. Claims for the
detection of the WHIM through soft X--ray emission have already been
presented \citep[e.g.,][]{zappacosta05,werner08}. However, these
detections are limited to special places, such as external regions of
massive galaxy clusters or large filaments connecting such clusters,
where the gas density reaches values higher than expected for the bulk
of the WHIM. A more promising approach to reveal the presence of the
WHIM lies in the detection of absorption features in the spectra of
background sources, at far ultraviolet (FUV) and soft X--ray energies,
associated to atomic transitions from highly ionized elements, using
GRB (Gamma Ray Burst) spectra \citep{fiore00,branchini09}, or in the
detection of Ly-$\alpha$ absorption from the tiny fraction of neutral hydrogen
revealed through FUV spectroscopy \citep[see][ for a
review]{richter08}. Although detection { of gas with} $T<10^5$K has
been obtained at FUV frequencies along a fairly large number of
sightlines \citep[e.g.,][ and references therein]{tripp08}, the
situation is less clear in the soft X--ray band. At these energies,
claims of WHIM detection in absorption have been presented by
different authors \citep[e.g.][]{nicastro05,buote09}, although they
are either controversial \citep[e.g.][]{kaastra06,rasmussen07} or have
moderate statistical significance.

Owing to the above observational difficulties in detecting and
characterizing the physical properties of the WHIM, numerical
simulations have played over the last years the twofold role to
forecast its detectability, both in emission
\citep[e.g.,][]{roncarelli06,ursino06} and in absorption
\citep[e.g.,][]{Cen_etal01,kravtsov02,viel03whim,chen03,viel05whim},
and to predict its observational properties, also as a function of the
physical processes included \citep[see][ for a review]{bertone08}. For
instance, feedback effects from star formation and accretion onto
super-massive black holes (SMBHs) are expected to determine at the
same time the process of galaxy formation and the physical properties
of the diffuse cosmic baryons. However, as of today, the mechanisms
powering feedback are still poorly understood and thus very difficult
to model in a fully self-consistent way in cosmological hydrodynamical
simulations. On the other hand, numerical simulations from different
groups
\citep[e.g.,][]{cenostriker06,dave07,koba07,oppe08,tvtb08,wiersma_etal09}
confirm that the thermo- and chemo-dynamical properties of the WHIM
and, more generally, of the Intergalactic Medium (IGM) at different redshifts are sensitive
to the nature and timing of the feedback.  Indeed, metals are observed
in the low density IGM out to high redshift and it is likely that
galactic winds, as observed in galaxies in the low redshift universe,
are responsible for the IGM metal enrichment. Furthermore, SMBHs are
also expected to have an impact on the gas distribution in galaxies
and to blow ejecta out to large distances, especially during galaxy
mergers.

In this paper, we will present an analysis of an extended set of
cosmological hydrodynamical simulations with the purpose of
investigating the effect that both galactic winds powered by supenova
(SN) explosions and energy feedback from accretion onto SMBHs have in
determining the thermal and chemical properties of the IGM. Although
the main focus of the analysis is on the low-redshift IGM, we will
also discuss how different feedback mechanisms leave their imprint on
the evolutionary properties of diffuse baryons since $z\sim 4$. Our
simulations are based on the chemo-dynamical version of the GADGET-2
code \citep{springel} presented by \cite{T07}, which follows the
production of different chemical species by accounting for detailed
yields from Type-Ia and Type-II SN (SN-Ia and SN-II hereafter), as
well as from intermediate and low mass stars in the
thermally-pulsating asymptotic giant branch (TP-AGB) phase, while also
accounting for the mass dependent life--times with which different
stellar populations release their nucleosynthetic products. The model
of galactic winds is that introduced by \cite{springel2003}, that we
consider both in its original version and in a version based on
assuming that winds are never hydrodynamically decoupled from the
surrounding medium \citep[see also][]{DallaVecchia_Schaye08}. As for
the BH feedback, we adopt the models originally introduced by
\cite{dimatteo05,springel05bh}. Besides investigating the effect of
different feedback mechanisms, we will also consider the impact of
changing the stellar initial mass function (IMF) on the resulting
enrichment pattern of the IGM. Although the results of this paper have
important implications for the detectability of the WHIM in view of
next-generation X--ray missions, we defer an observationally oriented
analysis of our simulations to a future work, aimed at investigating
in detail how different instrumental capabilities will be able to
characterize the WHIM properties at the level required to discriminate
between different feedback models.

The plan of the paper is as follows. In Section \ref{sim}, we describe
the hydrodynamical simulations used and the feedback mechanisms
adopted. Section \ref{results} describes the main results of our
analysis in terms of global gas properties (evolution of different
phases over redshift), epoch of enrichment of gas particles,
metallicity distributions at $z=0$ as a function of overdensity and
temperatures, and evolution of different chemical elements. We provide
a final discussion of our results and draw our main conclusions in
Section \ref{conclusions}. In the following, we will assume the values
of the solar metallicity as reported by \cite{asplund05}, { with
  $Z/X=0.0165$ ($X$: hydrogen mass; $Z$: mass contributed by all
  elements heavier than Helium)}.

\section{Hydrodynamical simulations}
\label{sim}
Our simulations were carried out using a version of the parallel
hydrodynamical TreePM-SPH code {\small {GADGET-2}} \citep{springel},
which includes a detailed implementation of the chemical enrichment as
described in \cite{T07}. The initial conditions are generated at
redshift $z=99$ in a cosmological volume with periodic boundary
conditions filled with an initially equal number of dark matter and
gas particles. The matter power spectrum is generated using CMBFAST
\citep{seljak1996} for a flat $\Lambda$CDM model, with cosmological
parameters consistent with the recent findings of WMAP year-5
\citep{komatsu09}: $\Omega_{\rm 0m }=0.24,\ \Omega_{0\Lambda}=0.76,\
\Omega_{\rm 0b }=0.0413$, for the density parameters contributed by
total matter, cosmological constant and baryons, $n_s=0.95$ for the
primordial spectral index, $H_0 = 73$ km s$^{-1}$ Mpc$^{-1}$ and
$\sigma_8=0.8$ for the normalization of the power spectrum. Besides
performing simulations within boxes of 37.5 $h^{-1}$ comoving Mpc
using $2\times 256^3$ gas and dark matter (DM) particles, we also run
a few simulations within a twice as large box of $75\hm$ on a side,
with $2\times 512^3$ particles so as to keep resolution constant. Such
larger boxes will allow us to keep under control any effect related to
box size. The resulting mass of the gas particles is thus $m_{\rm
  gas}\simeq 3.6\times 10^7\msun$. In these runs, the gravitational
softening is set to $\epsilon=7.5\, h^{-1}$ comoving kpc above $z=2$,
while at $z<2$ it is set to $\epsilon=2.5\, h^{-1}$ physical kpc. As
for the B-spline softening length used for the computation of the SPH
forces, the lowest allowed value was set to half of the gravitational
softening. In order to address resolution effects, we also carried out
one simulation in the smaller box using $2\times 400^3$ particles. In
this case the mass of the gas particles is $m_{\rm gas}\simeq
9.4\times 10^6\msun$, with all the softening lengths rescaled
according to $m_{\rm gas}^{-1/3}$: { $\epsilon=4.8\, h^{-1}$
  comoving kpc at $z>2$ and $\epsilon=1.6\, h^{-1}$ physical kpc at
  lower redshift}. All these simulations are evolved to $z=0$.

We point out that both a large volume and high resolution are
necessary in order to correctly describe the physical properties of
the WHIM. Indeed, shocks driven by the collapse of large-scale
structures are important, while dense environments able to trigger and
power the feedback processes need to be resolved. Our simulations of
different resolutions and/or box-sizes will be denoted by the tuple
(linear size of the box in comoving $\hm$, number of DM
particles$^{1/3}$). For instance, we will indicate with (75,512) a
simulation within a box of $75\hm$ and containing $512^3$ DM particles
(see Table 1).

Radiative cooling and heating processes are followed for a primordial
mix of hydrogen and helium and we include the dependence of cooling on
metallicity by adopting the cooling rates from the tables by
\cite{SD93}. We also include the effect of a spatially uniform
redshift--dependent Ultra Violet Background (UVB) produced by quasars,
as given by \citet{haardt1996}, with helium heating rates multiplied by
a factor 3.3 in order to better fit observational constraints on the
temperature evolution of the IGM. This background gives naturally a
hydrogen ionization rate $\Gamma_{-12}\sim 0.8$ at $z=2-4$
\citep{bolt05}, which is in broad agreement with the observations. We
adopt the effective model of star formation from a multiphase
interstellar medium introduced by \cite{springel2003}. In this model,
each gas particle whose density exceeds a limiting threshold value is
assumed to contain a hot and a cold phase, the latter providing the
reservoir of star formation. The two phases coexist in pressure
equilibrium with their relative fractions being computed according to
the local conditions of density and temperature. In our simulations we
assume that the density threshold for a gas particle to become
multiphase and, therefore, star forming, is $n_{\rm H}=0.1$
cm$^{-3}$ in terms of the number density of hydrogen atoms. Star forming gas
particles are then assumed to spawn collisionless gas particles
according to the stochastic scheme originally introduced by
\cite{KWH}. We allow each gas particle to produce up to three
generations of star particles, each having therefore a typical mass of
about one third of the initial mass of the gas particles.
  
Our simulations have been carried out with the chemo-dynamical version
of the GADGET-2 code described by \cite{T07}. The included model of
chemical evolution allows us to follow the production of six different
metal species (C, O, Mg, S, Si, Fe) from Type-II and Type Ia
supernovae (SNII, SNIa), along with low and intermediate mass stars
(LIMS) in the thermally-pulsating asymptotic giant branch (TP-AGB)
phase. Besides including different contributions from SNII, SNIa and
LIMS, we also include the effect of the mass--dependent time delay
with which different stellar populations release metals. Specifically,
we adopt the lifetime function given by \cite{padovanimatteucci93}. We
use the stellar yields by \cite{thielemann03} for SNIa, by
\cite{woosleyweaver95} for SNII and by \cite{vandenhoek97} for LIMS.
The mass-range for the SNII is $M > 8 M_{\rm \odot}$, while for SNIa
arising from binary systems in the mass range it is $0.8< M/M_\odot<
8$, with a binary fraction of 10 per cent. Finally, we use three
distinct stellar initial mass functions (IMFs): a Salpeter (1955), a
Kroupa (2001) and an Arimoto-Yoshii (1987) IMF. Our reference choice
is the functional form proposed by Kroupa (2001) which adopts a
multi-slope approximation, $\varphi(m) \propto m^{\rm -y}$ with
$y=0.3$ for stellar mass $m < 0.5\,M_{\rm \odot}$, $y=1.2$ for
$0.5\,M_{\rm \odot} \leq m < 1\,M_{\rm \odot}$ and $y=1.7$ for $m \geq
1\,M_{\rm \odot}$. Both the IMF by \cite{salpeter55} and that by
\cite{ay87} have instead a single slope, with $y=1.35$ and $y=0.95$,
respectively. Our model of chemical evolution also includes stellar
mass losses, which are self--consistently computed for a given IMF and
life-time function. This means that a fraction of the mass of the star
particles is restored as diffuse gas during the evolution and
distributed to the surrounding gas particles. We refer to \cite{T07}
for a detailed description of our implementation of chemo-dynamics in
GADGET-2, to \cite{saro06} and \cite{fabjan07} for applications to
simulations of galaxy clusters and to \cite{tvtb08} for a description
of the global IGM properties around DLAs at $z>2$.

In the two following subsections we will summarize the main features of
the two different feedback schemes adopted in this paper.

\subsection{Galactic outflows}

The implementation of galactic outflows is extensively described by
\cite{springel2003}.  In this model, winds are assumed to blow from
star forming regions with a mass loading rate $\dot{M}_{\rm w}$
proportional to the star formation rate $\dot{M}_{\rm \star}$
according to $\dot{M}_{\rm w}= \eta \dot{M}_{\rm \star}$.
Star-forming gas particles are then stochastically selected to become
part of a blowing wind, with a probability which is proportional to
their star formation rate. In its original implementation, whenever a
particle is uploaded to the wind, it is decoupled from hydrodynamics
until the density of the surrounding gas drops below a given limiting
value. This allows the wind particle to travel `freely' up to few kpc
until it has left the dense star-forming phase, without directly
affecting it. As a protection against decoupling a wind particle
indefinitely in case it gets 'stuck' in the ISM of a very massive
galaxy, the maximum allowed time for a wind particle to stay
hydrodynamically decoupled is set to $t_{\rm dec} = l_{\rm w} / v_{\rm
  w}$, where we fix $l_{\rm w}=10\,h^{-1}$kpc in our reference case,
while $v_{\rm w}$ is the wind speed. As for the limiting density for
hydrodynamic decoupling of gas particles, it is set to 0.5 in units of
the threshold for star formation. Unlike in \citet{springel2003}, we
decide here to fix the velocity of the winds to the value $v_{\rm
  w}=500\vel$, instead of fixing the fraction of the SNII energy
powering galactic ejecta. For the efficiency of the wind mass loading,
we adopt $\eta=2$. In sum, four parameters fully specify the wind
model: the wind efficiency $\eta$, the wind speed $v_{\rm w}$, the
wind free travel length $l_{\rm w}$ (whose detailed value is
unimportant) and the wind free travel density factor.

In order to verify the effect of the hydrodynamic decoupling of wind
particles we also perform a simulation without such a decoupling,
$t_{\rm dec}=0$, i.e. keeping the particles always hydrodynamically
coupled.  We note that \citet{nagamine07} showed that global
DLAs properties are relatively insensitive to the value of $l_{\rm
  w}$. On the other hand, \cite{DallaVecchia_Schaye08} pointed out
that keeping winds always hydrodynamically coupled has a significant
effect on the evolution and star formation in simulations of isolated
disk galaxies.

We stress that our wind model is not the only possible wind implementation and
others could be adopted such as those based on momentum--driven winds 
as suggested by \cite{murray05} and \cite{dave07}.

\subsection{BH feedback}
We also include in our simulations the effect of feedback energy from
gas accretion onto super-massive black holes (BHs), following the
scheme originally introduced by (\citealt{springel05bh}, see also
\citealt{dimatteo05}). In this model, BHs are represented by
collisionless sink particles initially seeded in just resolved DM
haloes, which subsequently grow via gas accretion and through mergers
with other BHs during close encounters. Every new dark matter halo,
identified by a run-time friends-of-friends algorithm, above the mass
threshold $M_{\rm th}=10^{10}\msun$, is seeded with a central BH
of initial mass $10^5 \msun$, provided the halo does not contain any
BH yet. Each BH can then grow by local gas accretion,
with a rate given by
\be
\dot M_{\rm BH}\,=\,{\rm min}\left(\dot M_{\rm B}, \dot M_{\rm Edd}\right)\,.
\label{eq:acrate}
\ee 
Here $\dot M_{\rm B}$ is the accretion rate estimated with the
Bondi-Hoyle-Lyttleton formula \citep[e.g.,][]{bondi52}, while $\dot
M_{\rm Edd}$ is the Eddington rate.  The latter is inversely
proportional to the radiative efficiency $\epsilon_{\rm r}$, which
gives the radiated energy in units of the energy associated to the
accreted mass: $\epsilon_{\rm r}=L_{\rm r}/(\dot M_{\rm BH}
c^2)$. Following \cite{springel05bh}, we use $\epsilon_{\rm r}=0.1$ as
a reference value, which is typical for a radiatively efficient
accretion onto a Schwarzschild BH \citep{shakura73}. The model then
assumes that a fraction $\epsilon_{\rm f}$ of the radiated energy is
thermally coupled to the surrounding gas, so that $\dot E_{\rm
  feed}=\epsilon_{\rm r} \epsilon_{\rm f} \dot M_{\rm BH}c^2$ is the
rate of the energy released to heat the surrounding gas. Using
$\epsilon_{\rm f}\sim 0.05$, \cite{dimatteo05} were able to reproduce
the observed $M_{\rm BH}-\sigma$ relation between bulge velocity
dispersion and mass of the hosted BH (see also
\cite{sija08,dimatteo08}).  Gas particle accretion onto the BH is implemented
in a stochastic way, by assigning to each neighbouring gas particle a
probability of contributing to the accretion, which is proportional to
the SPH kernel weight computed at the particle position.  In the
scheme described above, this stochastic accretion is used only to
increase the dynamic mass of the BHs, while their mass entering in the
computation of the accretion rate is followed in a continuous way, by
integrating the analytic expression for $\dot M_{\rm BH}$. Once the amount
of energy to be thermalised is computed for each BH at a given
time-step, this energy is then distributed to the surrounding gas
particles using the SPH kernel weighting scheme.

\vspace{0.3truecm} 

To sum up, we run simulations with the \cite{Kroupa01} IMF by turning
off galactic winds (NW), with galactic winds (W), with galactic winds
always hydrodynamically coupled (CW) and with black hole feedback
(BH). We summarize in Table 1 the simulations analysed in
this paper. Overall, our simulation set allows us to address the
effect of changing: {\em (a)} box size (by comparing W$_{37,256}$ with
W$_{75,512}$ and Way$_{37,256}$ with Way$_{75,512}$), {\em (b)} IMF
(by comparing W$_{37,256}$ with Way$_{37,256}$ and with
Ws$_{37,256}$), {\em (c)} resolution (by comparing W$_{37,256}$ with
W$_{37,400}$) and {\em (d)} nature of the energy feedback (by comparing
W$_{37,256}$ with NW$_{37,256}$, with CW$_{37,256}$ and with
BH$_{37,256}$).

\begin{table*}
\label{t:runs}
\begin{tabular}{lcccc}
\hline 
  Run & Box size & $N_{\rm DM}^{1/3}$ & IMF & Feedback 
 \\ \hline 
W$_{37,256}$ & 37.5 & 256 & Kroupa & Winds; $v_{\rm w}=500\vel$ \\ 
W$_{75,512}$ & 75.0 & 512 & Kroupa & Winds; $v_{\rm w}=500\vel$ \\ 
W$_{37,400}$ & 37.5 & 400 & Kroupa & Winds; $v_{\rm w}=500\vel$ \\ 
Way$_{37,256}$ & 37.5 & 256 & Arimoto-Yoshii & Winds; $v_{\rm w}=500\vel$ \\ 
Ws$_{37,256}$ & 37.5 & 256 & Salpeter & Winds; $v_{\rm w}=500\vel$ \\ 
NW$_{37,256}$ & 37.5 & 256 & Kroupa & No feedback \\ 
CW$_{37,256}$ & 37.5 & 256 & Kroupa & Coupled winds; $v_{\rm w}=500\vel$ \\ 
BH$_{37,256}$ & 37.5 & 256 & Kroupa & Black Hole feedb., no winds 
\\ \hline
\end{tabular}
\caption{Summary of the different runs. Column 1: run name; column 2:
  comoving box size (units of $\hm$); column 3: number of DM particles;
  column 4: stellar initial mass function (IMF, see text);
  column 5: feedback included (see text).}
\end{table*}

\section{Results}
\label{results}

\begin{figure}
\includegraphics[width=0.52\textwidth]{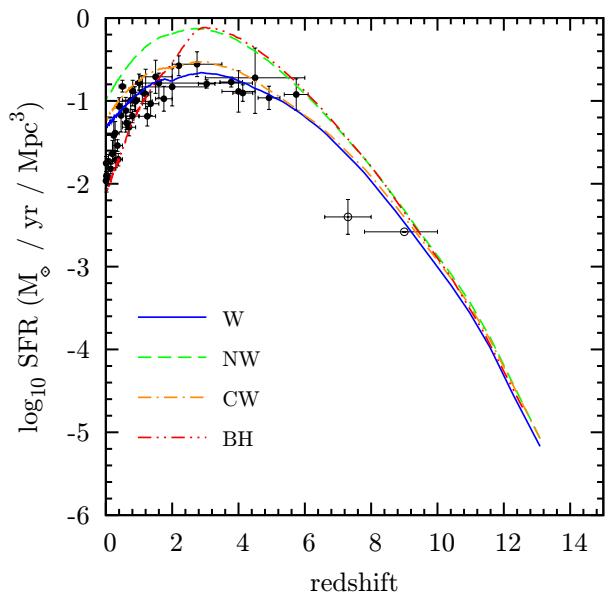}
\caption{Star formation rate density as a function of redshift in
  units of M$_{\odot}$yr$^{-1}$Mpc$^{-3}$ for W$_{37,256}$ (winds,
  continuous blue line), NW$_{37,256}$ (no feedback, dashed green
  line), CW$_{37,256}$ (coupled winds, dot-dashed orange line) and
  BH$_{37,256}$ (black hole feedback, triple dotted-dashed line) runs.
  Observational data points represented as full circles are taken from
  \protect\cite{hopkins04}, while the empty circles are from
  \protect\cite{bouwens08}.}
\label{figsfr}
\end{figure}

In Figure \ref{figsfr} we show the star formation rates (SFR) for the
W$_{37,256}$, NW$_{37,256}$, CW$_{37,256}$ and BH$_{37,256}$
simulations along with observational results by \cite{hopkins04}. At
$z\magcir 3$ the NW and BH runs behave very similarly. This is not
surprising since BH feedback is not effective until a sufficiently
large number of DM haloes, massive enough to host a seed BH, is
numerically resolved. After that, gas accretion takes place in BHs
with a subsequent release of thermal energy.  Once BH feedback becomes
efficient, star formation is suddenly suppressed by the expulsion of
hot gas. Kinetic feedback by galactic winds is relatively more
efficient at high redshifts in depriving relatively small haloes of
the star--forming gas, while it fails in regulating the star formation
within massive haloes at later epochs. We note that hydrodynamically
coupled winds (CW) tend to produce a slightly higher SFR around the
peak of the star formation history. This is due to the fact that
coupled winds tend to thermalize their kinetic energy at smaller
distances from star forming and dense regions. The cooling time of
this gas is correspondingly shorter, which causes a larger fraction of
the thermalized wind energy to be radiated away, thus reducing the
feedback efficiency. Both runs including winds reproduce the observed
behaviour of the SFR at high redshift, $z\magcir 2$, while they tend
to produce too high star formation at low redshift. On the contrary,
the run with BH feedback has a too high star formation at $z>3$, while
it recovers the observed SF level at $z\mincir 2$. This illustrates
once again the different role played by feedback related to star
formation and to BH accretion: while the former is efficient since
early times in regulating gas cooling within small haloes, the latter
sets in at relatively lower redshift to quench star formation within
recently formed massive haloes.

\subsection{Global gas properties}
\label{globalgasproperties}

\begin{figure*}
\includegraphics[width=1\textwidth]{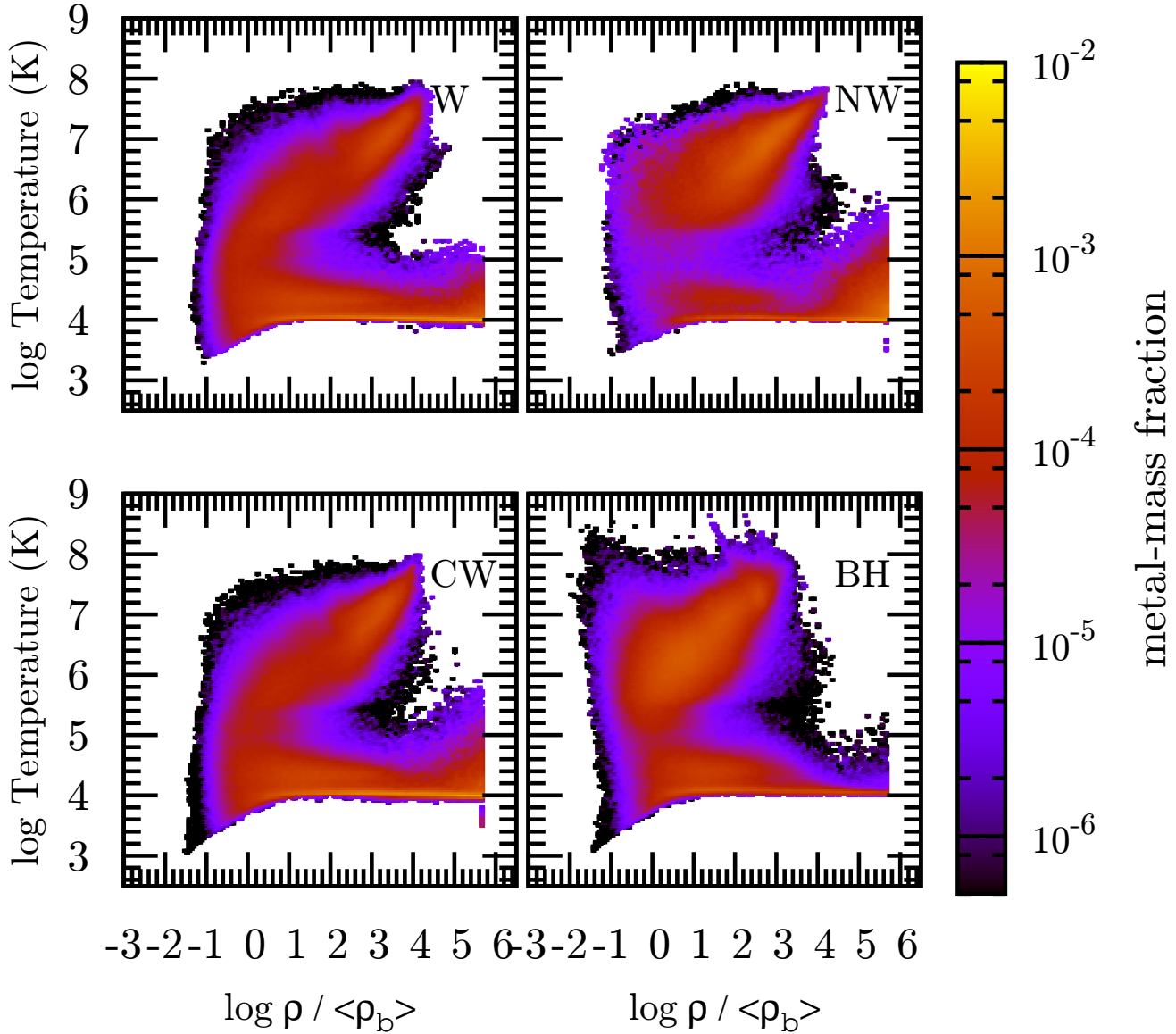}
\caption{The $\rho_{\rm gas}-T$ phase diagrams at $z=0$ for
  simulations based on the \protect\cite{Kroupa01} IMF, by varying the
  feedback scheme, color coded according to the metal mass fraction
  (see vertical bar): NW$_{37,256}$ run with no feedback (top left),
  W$_{37,256}$ run with galactic winds (top right), CW$_{37,256}$ run
  with galactic winds always hydrodynamically coupled (bottom left) and
  BH$_{37,256}$ run with BH feedback (bottom right).}
\label{fig_phasespace}
\end{figure*}

We show in Figure \ref{fig_phasespace} the density--temperature phase
diagrams for the gas in the four simulations based on the
\cite{Kroupa01} IMF and including different feedback schemes. The four
panels are for the simulations without feedback (NW$_{37,256}$, top
left), with feedback associated to galactic winds (W$_{37,256}$, top
right), with galactic winds always hydrodynamically coupled
(CW$_{37,256}$, bottom left) and with BH feedback (BH$_{37,256}$,
bottom right). Each phase diagram is color-coded according to the mass
in metals which is associated to the gas particles belonging to each
two-dimensional bin in the $\rho_{\rm gas}$--$T$ plane, with brighter
regions corresponding to a larger metal mass. As such, these plots
convey information on the effect that different feedback schemes have
on the way in which metals are distributed in density and temperature.
{ Here and in the following we do not include in the baryon budget
  inventory the inter--stellar medium (ISM), i.e. for gas having
  density larger than the density threshold for the star formation
  (see Sect. 2). We note that only a tiny fraction of the gas lies in
  this phase at all redshifts. Furthermore, while our sub-grid model
  for star formation should provide an effective description of the
  ISM, it is not obvious that this description provides a fully
  realistic description of its temperature. For this reason, our
  gas mass and metal budget will not include the contribution from
  star--forming gas.}

The effect of feedback is quite strong for gas in the temperature
range $10^5$--$10^7$ K (corresponding to the WHIM phase that we will
define below; e.g. \citealt{cenostriker06}). Strong galactic winds
increase the amount of metals carried by gas with temperature in the
range $10^5-10^6$K and overdensity \footnote{Here and in the following
  we denote with $\db$ the overdensity of gas with respect to the mean
  cosmic baryon density $\bar\rho_{\rm b}$, i.e. $\db =\rho_{\rm
    b}/\bar\rho_{\rm b}-1$.} $\db\simeq 1$--10, when compared to the
no feedback case. This is a consequence of the fact that winds are
loaded with gas particles which were star forming and, as such, have
been heavily enriched. The effect is even more dramatic in the
presence of BH feedback: even regions below the mean density are
enriched with metals up to rather high temperatures ($\sim 10^7$ K)
when compared to the corresponding W and CW simulations. { This
  counter-intuitive result follows from the fact that BHs' release of
  large amount of energy in a relatively short time interval, around
  $z\sim 3$. This sudden energy release turns out to be much more
  efficient than winds to heat metal enriched gas to high
  temperatures, thus displacing it from the haloes of star-forming
  regions to low-density regions. Once brought to high entropy by BH
  feedback, this enriched gas is prevented from re-accreting into
  collapsed haloes at lower redshift.}  Gas in photo-ionization
equilibrium at $T\sim 10^4$ K contains a comparable amount of metals
in the three runs which include feedback. In the NW case the amount of
metals present in the gas around the mean density is an order of
magnitude lower than in the runs including winds. Indeed, in the NW
simulation, most of the enriched gas remains at high density, instead
of being transported away from star-forming regions. The short cooling
time of this enriched gas causes its selective removal from the
diffuse phase into the stellar phase. We note also that relatively
cold gas in very dense regions, $\rho_{\rm gas}\simeq
(10^3-10^4)\bar\rho_{\rm b}$, is more enriched with metals in the NW,
CW and W simulations than in the BH one. This demonstrates the
efficiency of BH feedback in displacing highly enriched gas outside
the core regions of virialized haloes, while correspondingly
increasing the enrichment level of gas at lower density. As for the
comparison between coupled and decoupled winds (W and CW runs,
respectively), the latter has more enriched gas at densities
approaching the threshold for the onset of star formation.  This is
the consequence of the hydrodynamic coupling of winds which causes
wind particles to remain more confined in the proximity of
star--forming regions.

From a qualitative inspection of Fig. \ref{fig_phasespace} we draw the
following conclusions: i) galactic winds have a large impact on the
metal enrichment of the IGM at $z=0$ both for underdense-cold and mean
density-hot regions of the $T-\rho_{\rm gas}$ plane; ii) AGN feedback
is more efficient than winds in enriching the warm-hot
gas at $T=10^5-10^7$ K and relatively low overdensity, $\db\mincir
50$; iii) correspondingly, very dense regions, with overdensity
$\db>10^4$, are less enriched with metals in the BH simulations than
in the other runs. 

\subsection{Redshift evolution of the different phases} 

\begin{figure*}
\includegraphics[width=0.33\textwidth]{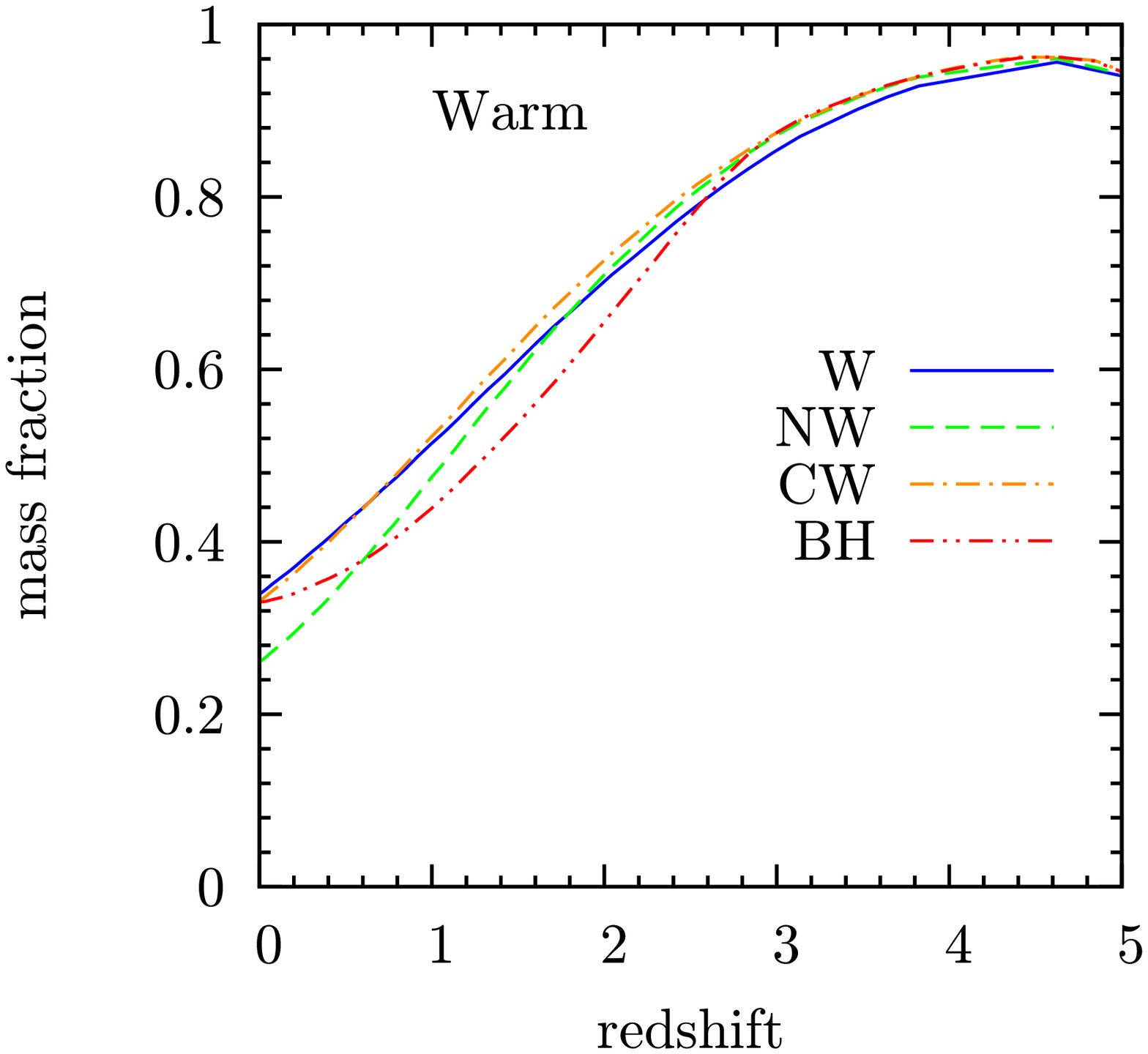}\includegraphics[width=0.33\textwidth]{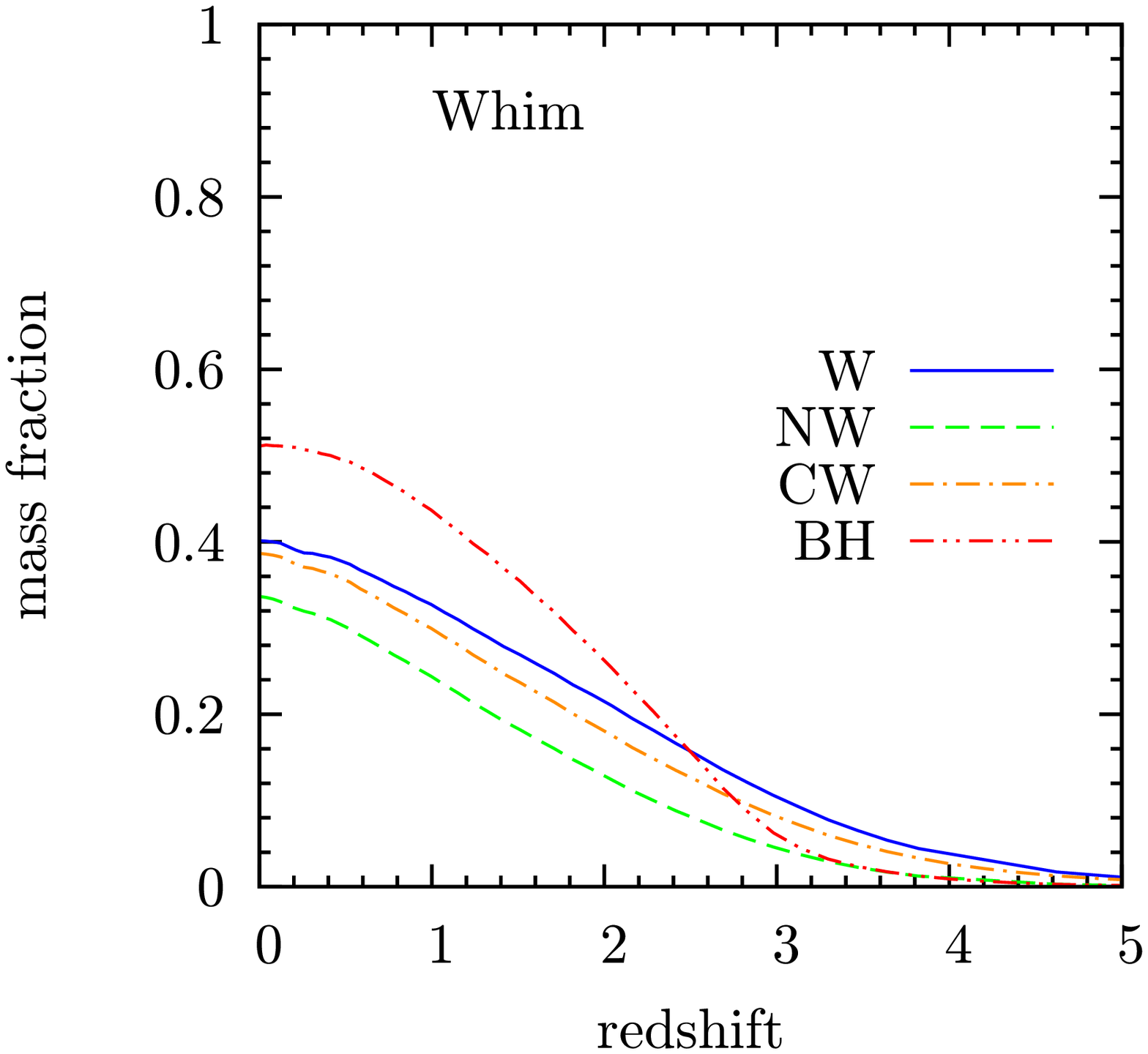}\includegraphics[width=0.33\textwidth]{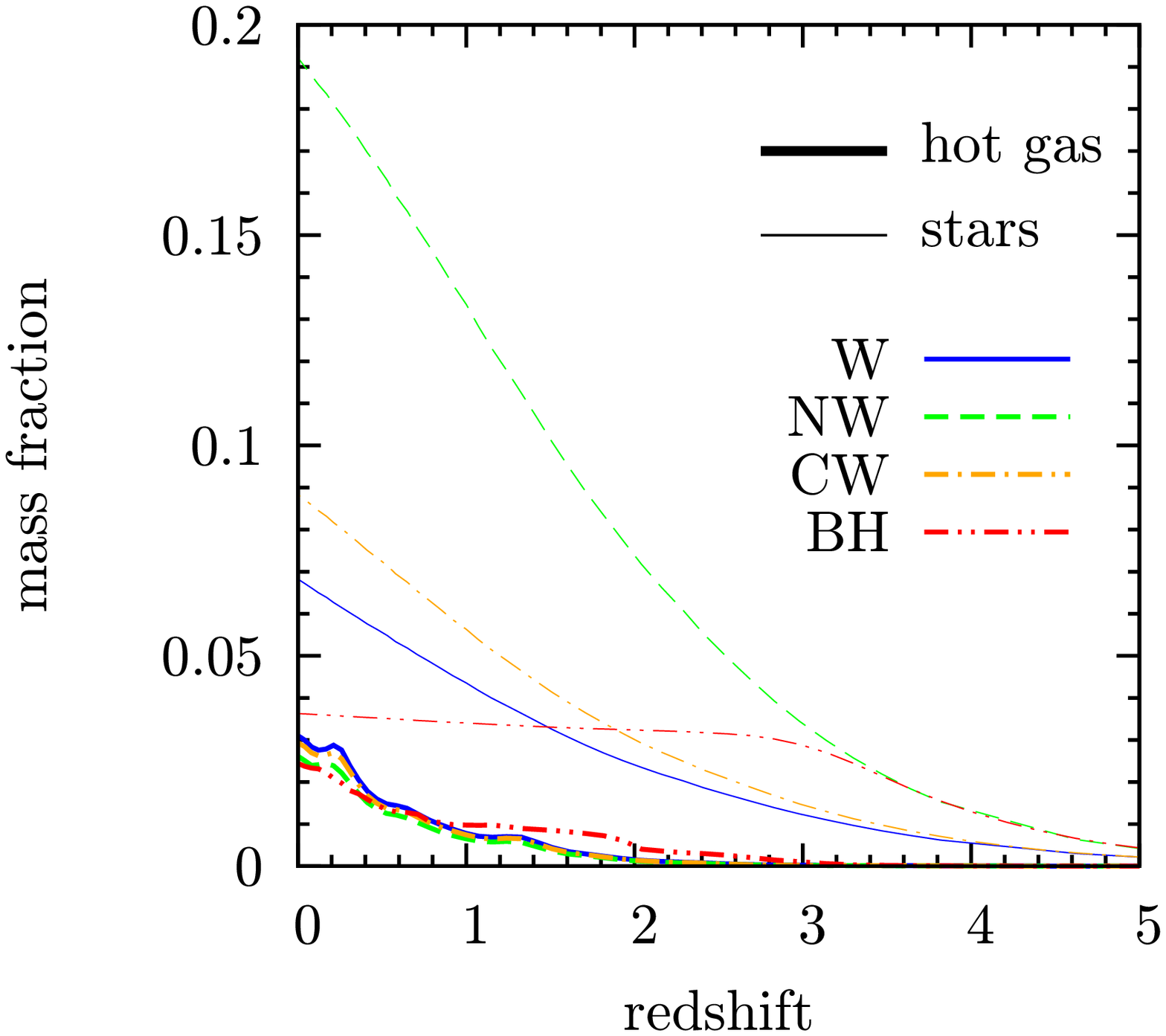}
\caption{The fraction of mass in the warm (left panel), WHIM (middle
  panel) and stellar plus hot phases (right panel, thin lines for
  stars and thick lines for hot gas) as a function of redshift for the
  NW$_{37,256}$, W$_{37,256}$, CW$_{37,256}$ and BH$_{37,256}$ runs
  (continuous blue, dashed green, dot-dashed orange and double-dot
  dashed red curves, respectively).}
\label{fig_massfraction}
\end{figure*}

\begin{figure*}
\includegraphics[width=0.33\textwidth]{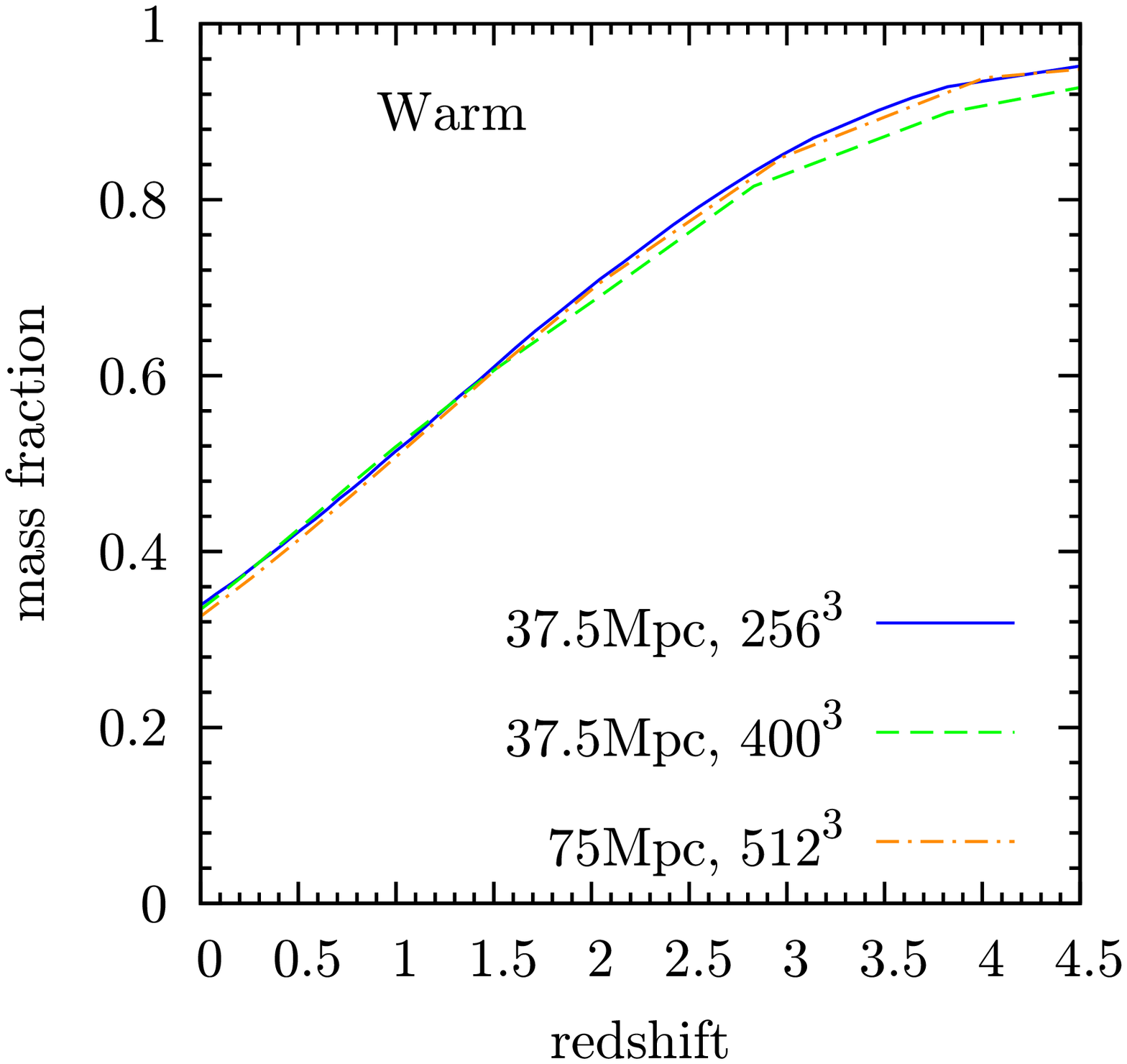}\includegraphics[width=0.33\textwidth]{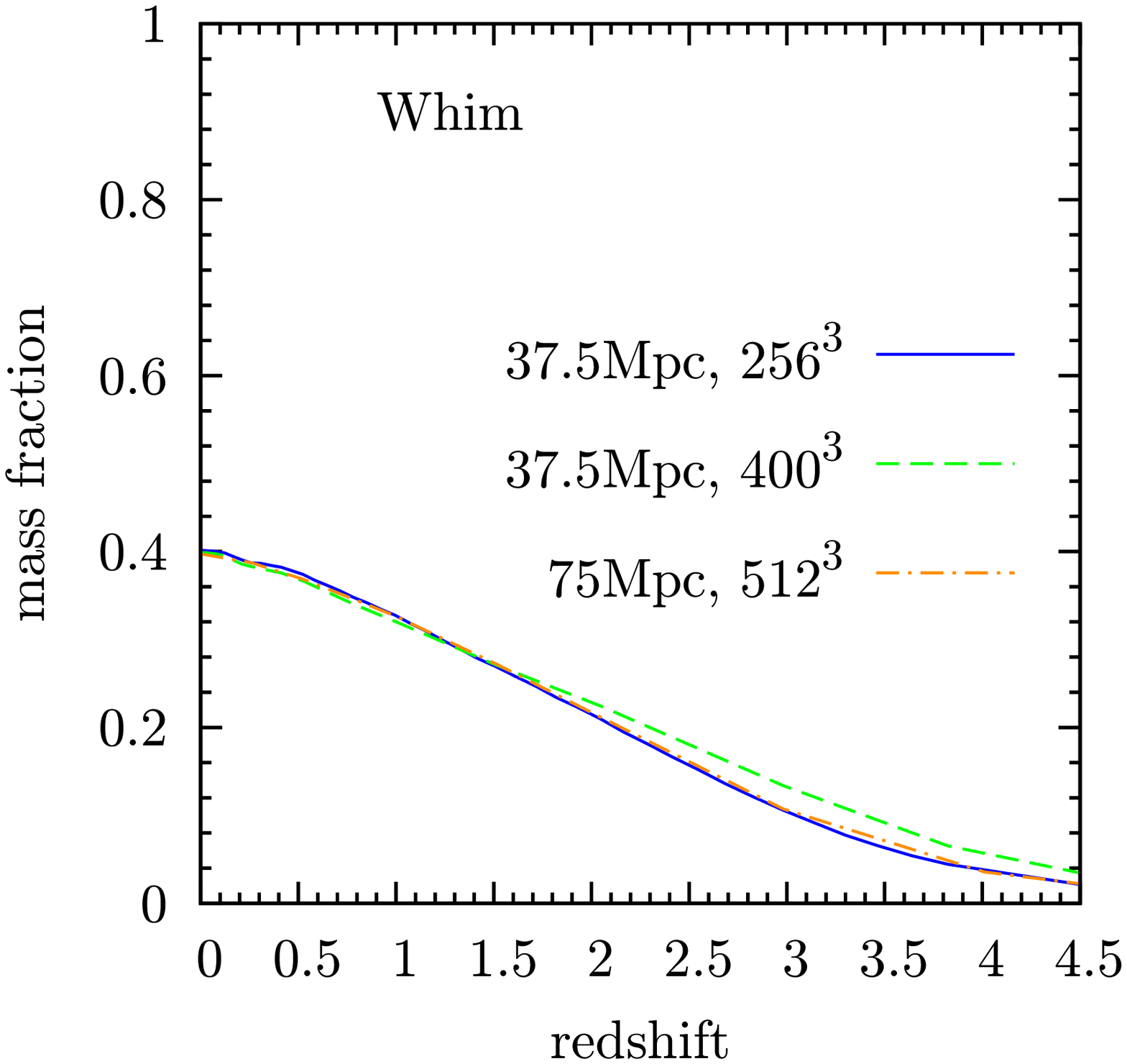}\includegraphics[width=0.33\textwidth]{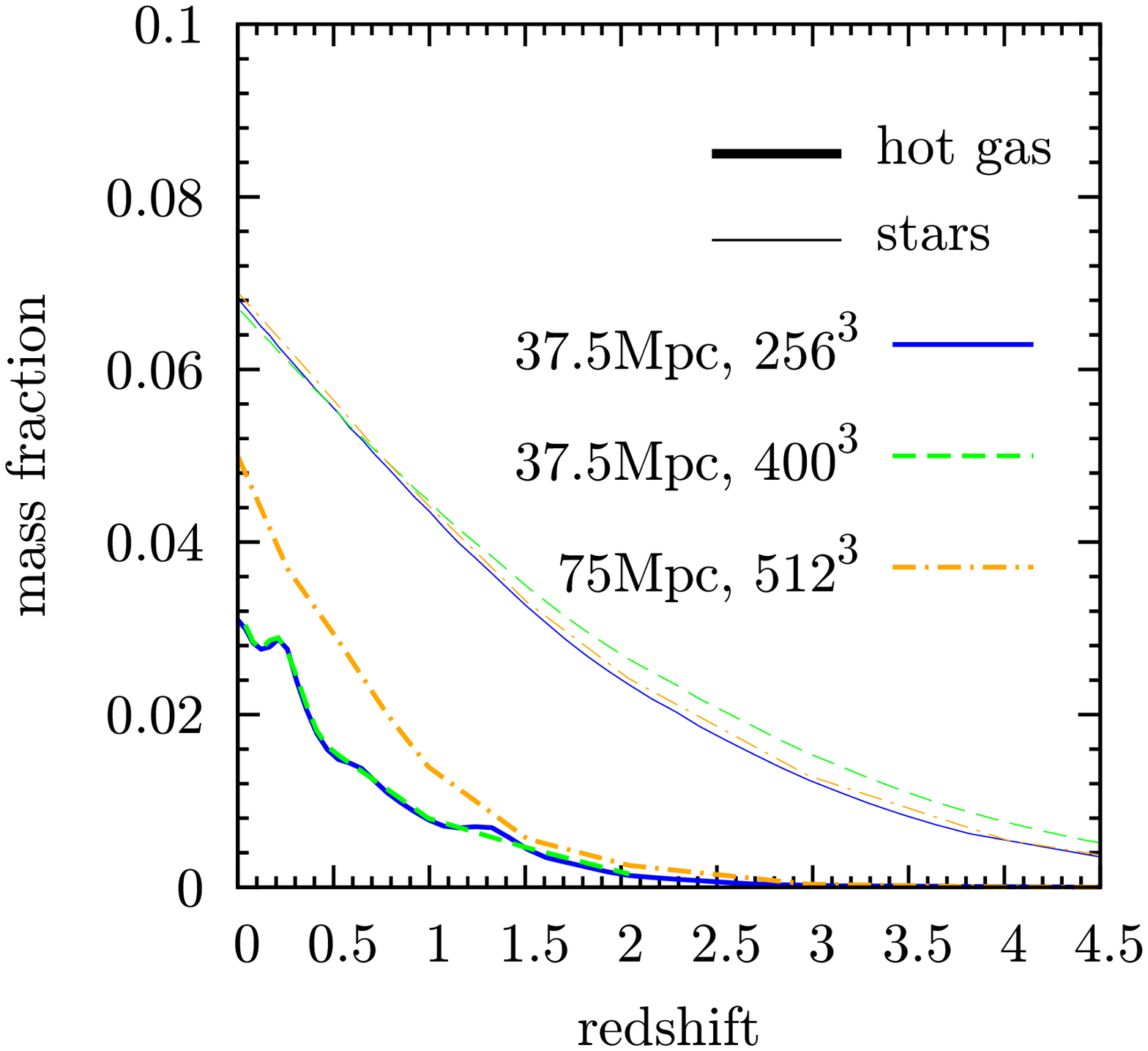}
\caption{Effects of resolution and box-size on the amount of gas in
  different phases. The fraction of mass in the warm (left panel),
  WHIM (middle panel) and stars plus hot components (right panel, thin
  lines for stars and thick lines for hot gas) as a function of
  redshift for the W$_{37,256}$, W$_{37,400}$, W$_{75,512}$ runs
  (continuous blue, dashed green, dot-dashed orange curves,
  respectively).}
\label{fig_massfraction_resolution}
\end{figure*}

In order to make our results directly comparable with those from
previous analyses \citep[e.g.,][]{dave01,cenostriker06}, we define
four different phases for the baryons in our simulations: a WHIM
phase, for gas particles with temperature in the range $10^5-10^7$ K;
a warm phase, for gas particles with $T=10^4-10^5$ K; a hot phase, for
gas particles with $T>10^7$ K; a condensed phase, which includes all
the baryonic mass associated with star particles.

In Figure \ref{fig_massfraction}, we show the redshift evolution of
{ the mass fraction in} these four different phases, comparing the
effect of changing feedback scheme. The W$_{37,256}$, NW$_{37,256}$,
CW$_{37,256}$ and BH$_{37,256}$ runs are represented by the continuous
blue, dashed green, dot-dashed orange and double-dot dashed red lines,
respectively. The warm phase is reported in the left panel. Although
showing slightly different evolutions in detail, the amount of warm
gas at $z=0$ is almost the same, $\simeq 35$ per cent, for the runs
with galactic winds and with BH feedback. This gas should be partly
responsible for local \lya forest absorption
\citep[e.g.,][]{danforth08}, samples the outskirts of galactic haloes
and is weakly sensitive to the energetics and mass-loading parameters
of the energy-driven galactic wind feedback \citep{pbv08}. The case
with no feedback (NW) has instead a $\sim 10$ per cent lower amount of
warm gas at $z=0$. In the middle panel, we show the evolution of the
WHIM phase: with BH feedback the amount of WHIM at $z=0$ reaches 50
per cent, about 10 per cent more than for the runs with winds and 15
per cent higher than for the NW case. Similarly to the results for the
warm phase, there are small differences between the W and CW runs,
{ also in the amount of WHIM gas and stars. A more significant
  difference between W and CW runs is found for the mass fraction in
  stars. The reason for this difference is in the lower efficiency of
  the CW feedback model. Indeed, hydrodynamically coupled winds
  deposit kinetic energy through hydrodynamical processes in
  relatively higher-density environments, which are characterized by
  shorter cooling times, which causes the thermalized energy to be
  promptly radiated away.}

The evolution of the WHIM mass fraction in the BH run is initially
identical to that for the NW run at $z\magcir 3.5$. Indeed, at high
redshift there is still a limited number of DM haloes whose mass is
large enough to host actively accreting BHs. At lower redshift, when
accretion onto BHs is more effective, the released feedback energy
efficiently heats the gas surrounding galaxy--size haloes, thus moving
gas from the warm to the WHIM phase. This explains at the same time
the lower amount of warm gas in the BH run at intermediate redshift
and the correspondingly larger amount of WHIM. As for the effect of
galactic winds, they play a role in heating circum--galactic gas
already at high redshift, soon after the onset of star formation. This
explains the larger amount of WHIM at $z\magcir 2.5$ in the W and CW
runs with respect to the BH run. The situation is reversed at lower
redshift, when the depths of the forming potential wells become large,
making gas heating with winds less effective with respect to BH
feedback.

In the right panel we plot the evolution of the mass fractions in stars
and hot gas, which are represented by the thin and thick lines,
respectively (note the different scale in the $y-$axis). As expected,
the mass in stars in the runs with and without feedback is
significantly different, largely reflecting the behaviour of the SFR
histories shown in Fig. \ref{figsfr}: the BH run has a fraction of
stars which is five (two) times less than the NW (CW) run, as a
consequence of the quenching of star formation taking place around
$z=3$, while the W run produces a slightly smaller amount of stars
when compared to the CW one.

Recently, \cite{li_white09} used SDSS data to reconstruct the galaxy
stellar mass function in the local Universe. Assuming an IMF by
\cite{chabrier03} (very close to the IMF by \citealt{Kroupa01} used in
the simulations shown in Figure \ref{fig_massfraction}), they
concluded that 3.5 per cent of the baryonic mass is locked into stars.
{ While clearly ruling out a model with no feedback, this
  observational estimate of the stellar mass fraction is close to the
  predictions of the BH run, although the star formation rate in this
  case is too high at $3\mincir z \mincir 6$ and slightly too low at
  $z\mincir 0.5$. As for the W and CW runs, they produce about two
  times and 2.5 times more stars than the BH run, irrespectively, as a
  consequence of a star formation excess below $z\sim 1$. These
  results suggest that both feedback mechanisms should be active: SN
  driven galactic ejecta should regulate star formation already at
  high redshift within relatively small galaxies, while BHs should
  quench cooling in large haloes around the maximum of the star
  formation, thereby keeping gas pressurised within such haloes down
  to $z=0$.}

The amount of hot gas at temperatures above $10^7$K is instead very
similar for all the simulations and around 3 per cent. This is a
consequence of the fact that the bulk of hot gas lies within groups
and clusters of galaxies, whose gas content is marginally affected by
the details of the feedback. We note, however, that while the amount
of hot gas in the BH run is larger than in the other runs at high
redshift, it drops below them at $z=1$, with an overall trend that is
opposite to that of the WHIM component. Indeed, at high redshift BH
feedback is quite efficient in heating gas to high temperature, thus
providing a sort of diffuse pre--heating. At later time, this
pre--heated gas has a harder time to fall into group--size potential
wells. To quantify this effect, we note that in the BH run about $\sim
10$ per cent of the hot gas lies below $\db\, \sim 1$, while only
$\mincir 0.1$ per cent of hot gas lies at such low densities in the
other runs.

{ It is worth pointing out that although at $z=0$ the total amount
  of hot gas is about the same, $\simeq 3$ per cent, for all feedback
  models, its distribution as a function of density has a distinct
  pattern in the BH run.  In fact, in the runs with no BH feedback gas
  is heated mostly by the process of gravitational virialization
  within the potential wells of galaxy groups and clusters. As a
  consequence, in those runs only about the 5 per cent of hot gas lies
  below virial overdensities, $\db\sim 50$.  On the other hand, a
  sizeable amount of gas displaced by BH feedback lies outside
  virialised haloes, with $\sim 37$ per cent of it found at
  $\delta_b\mincir 50$. Furthermore, the heating efficiency of BH
  feedback prevents hot gas from reaching densities as high as in the
  other runs in the central regions of galaxy groups and clusters (see
  also \citealt{sija07,bhatta08,fabjan_etal09}). As a result, the
  highest density reached by hot gas in the BH runs is lower by about
  a factor three than in the NW, W and CW runs.

  Interestingly, hot metals follow the fate of the gas they are
  associated with. Indeed, the fractions of hot metal mass lying at
  different densities are quite similar to the corresponding fractions
  of hot gas mass.}

In order to verify the robustness of our results against box size and
numerical resolution, we compare in Figure
\ref{fig_massfraction_resolution} the results of the W$_{37,256}$ run
with those of the W$_{75,512}$ and W$_{37,400}$ runs. In general, we
find that there is a very good convergence in the values of the mass
fractions associated to the WHIM, warm and star phases, at all
considered redshifts. We only note that the amount of gas in the hot
phase significantly increases in the W$_{75,512}$ run at $z\mincir
1.5$. This is due to the fact that a larger box can accommodate a
larger number of relatively more massive galaxy groups and clusters
where a larger amount of gas is shock--heated to a temperature above
$10^7$K. { A larger box, with size $\sim 200\hm$, would certainly
  allow to better sample the high end of the halo mass function and,
  therefore, to provide a fully converged estimate of the mass
  fraction in hot gas. We note, however, that our larger simulation
  box already allow us to sample the scale of galaxy groups, which
  contain gas at a temperature $\magcir 10^7$ K. Indeed, the largest
  halo found within the 75$\hm$ box has a mass of about $3\times
  10^{14}\msun$ and a X--ray emission--weighted temperature of about
  $4\times 10^7$K. Therefore, we do not expect our baryon inventory in
  simulations to be significantly affected by effects of finite box
  size.} In the following, we will restrict our analysis to the four
(37, 256) runs, based on the four feedback schemes (NW, W, CW,
BH). For this reason, and unless otherwise specified, we omit from now
on the box-size and number of the particles when referring to the
analysed simulations.

In summary, we conclude that at $z=0$ the amount of mass in the WHIM
in the simulation including BH feedback is about 10 per cent larger
than in the simulations including galactic winds, and also displays a
different redshift evolution. Quite remarkably, this result is in
quantitative agreement with that reported by \cite{cenostriker06}. The
fact that comparable values for the WHIM mass fraction are found with
different simulation codes, based on different hydrodynamic schemes,
and using different implementations of feedback processes highlights
that this should be considered as a robust prediction of models of
cosmic structure formation. { Therefore, if future observations
  will falsify these predictions, this will have direct implications
  on the need to include new physical processes in simulations.}

\subsection{The history of enrichment}

\begin{figure}
\includegraphics[width=0.5\textwidth]{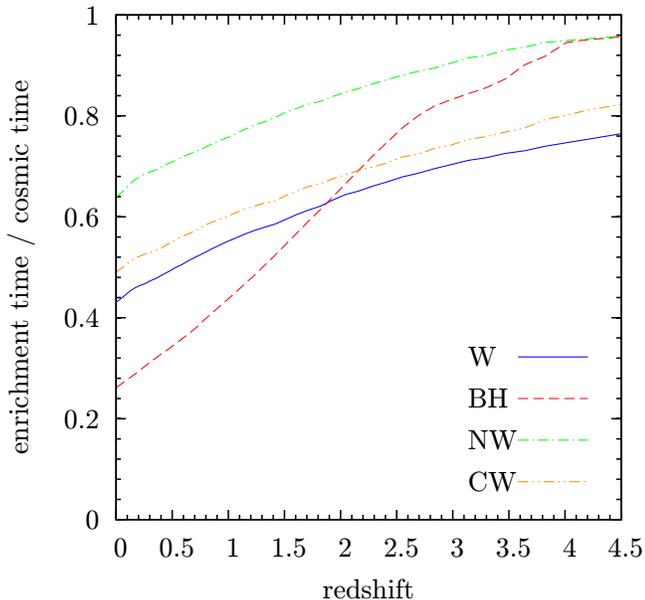}
\caption{Ratio between the average age of WHIM enrichment and the age
  of the universe as a function of redshift ($\bar{t}_{\rm
    WHIM}/t_{\rm cosmic}$). Results are shown for the W$_{37,256}$,
  NW$_{37,256}$, CW$_{37,256}$ and BH$_{37,256}$ simulations,
  represented by the continuous blue, dot-dashed green,
  double-dot-dashed orange and dashed red lines, respectively.}
\label{fig_epoch}
\end{figure}

\begin{figure*}
\includegraphics[width=0.5\textwidth]{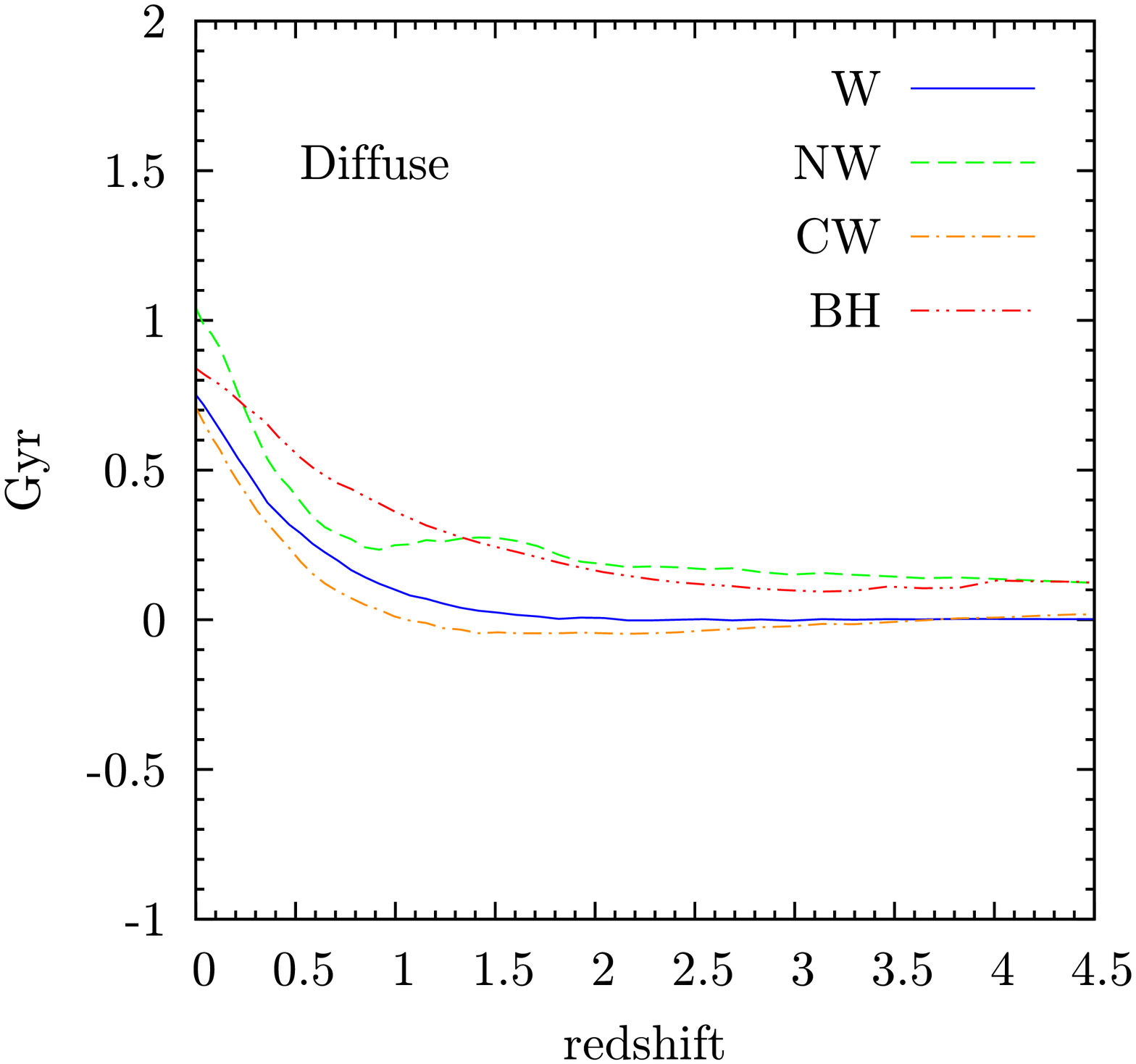}\includegraphics[width=0.5\textwidth]{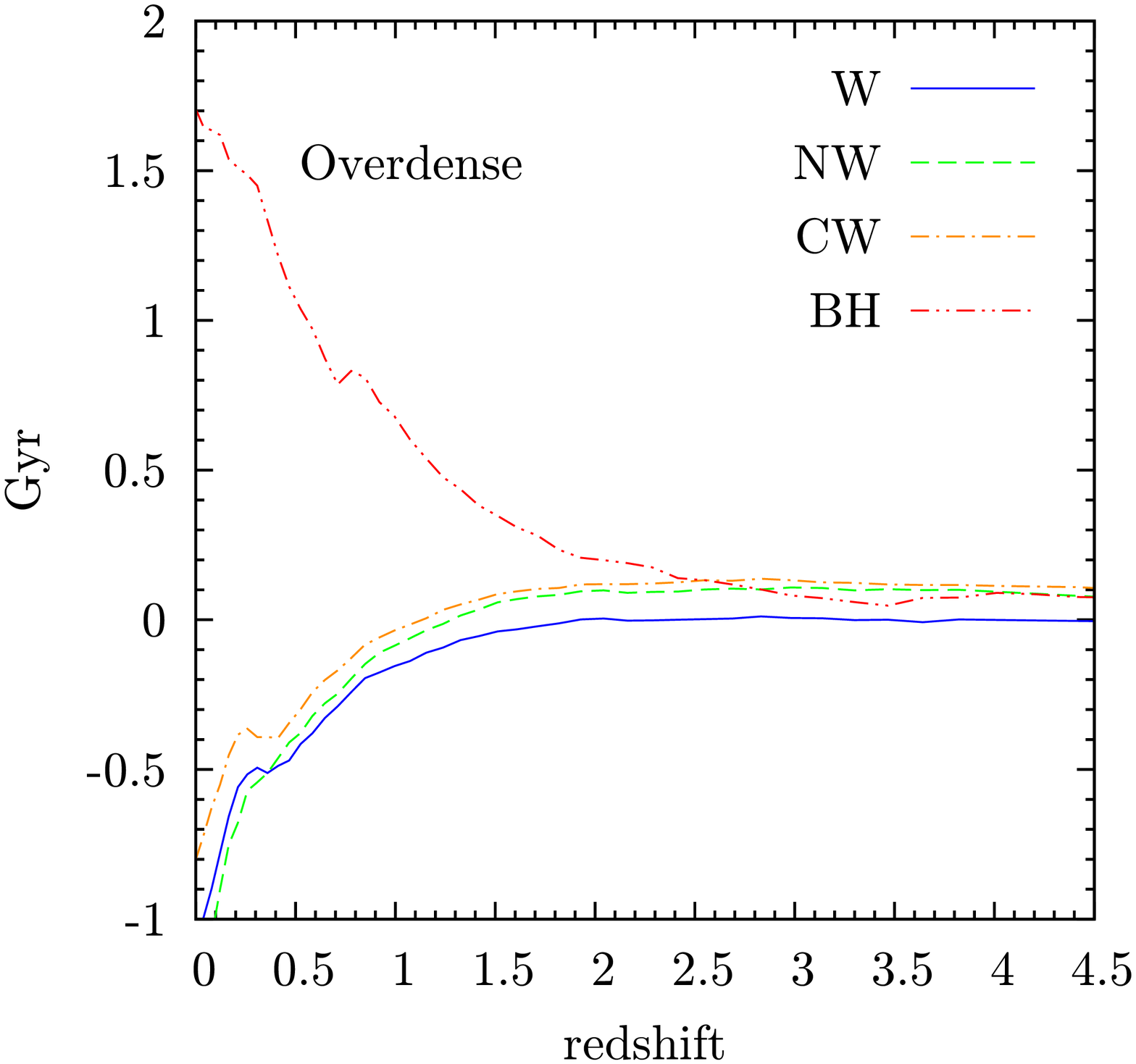}
\caption{Left panel: difference, in
  Gyr, between the average ages of enrichment for WHIM and warm
  particles in the diffuse phase ($\bar{t}_{\rm WHIM,\db\, < 50} -
  \bar{t}_{\rm warm,\db\, < 50}$). Right panel: the same as in the
  left panel but for gas in the collapsed phase ($\bar{t}_{\rm
    WHIM,\db\, > 50} - \bar{t}_{\rm warm,\db \, > 50}$). Results are
  shown for the W$_{37,256}$, NW$_{37,256}$, CW$_{37,256}$ and
  BH$_{37,256}$ simulations, represented by the continuous blue,
  dot-dashed green, double-dot-dashed orange and dashed red lines,
  respectively.}
\label{fig_epoch_diff}
\end{figure*}

In Figure \ref{fig_epoch} we show the redshift dependence of the
average age of enrichment of the gas particles in units of the age of
the universe at the same redshift. Results are shown for warm and
WHIM phases. Besides showing results for all gas particles belonging
to each of these two phases, in the following we will also classify gas
within each phase according to its overdensity. Specifically, we will
define a {\em collapsed} phase, which is made up by all gas particles having
overdensity $\db>50$ (we note that this is the typical overdensity
reached within virialised haloes), and a {\em diffuse} phase, which
contains gas at $\db<50$.

Furthermore, we define the average age of enrichment of a gas particle
at redshift $z$ as
\begin{equation}
\bar t(z)\,=\,{\sum_i \Delta m_{Z,i}(z) t_i\over m_{Z,i}(z)}\,,
\label{eq:avenr}
\end{equation}
where the sum is over all the time-steps performed until redshift $z$,
$\Delta m_{Z,i}$ is the mass in metals received by the particle at the
$i$-th simulation time-step, $t_i$ is the cosmic time at that
time-step and $m_Z(z)$ is the total metal mass received by a particle
before $z$. According to this definition, large values of enrichment
age, at a given redshift, correspond to a smaller look-back time for
the epoch of enrichment, while smaller values of $\bar t(z)$ indicate
a more pristine metal enrichment. In the limit in which all the metals
were received by the particles at the considered redshift $z$, then the
enrichment age coincides with the cosmic age at that redshift.  The
mean of the ages of enrichment of a given phase is then computed by
averaging eq. (\ref{eq:avenr}) over all particles belonging to that
phase, each particle being weighted according to its mass in metals:
\begin{equation}
  \bar t_{\rm phase}(z)\,=\,{\sum_j \bar t_j(z) m_{Z,j}(z)\over \sum_j m_{Z,j}(z)}\,.
\label{eq:avenr_particles_in_phase}
\end{equation}
It is worth pointing out that this way of computing the enrichment age
of a gas phase does not provide a measurement of the average time of
enrichment of the phase to which the particles belong; instead, it is
an estimate of the average time at which the metals that belong to
that phase were deposited in the gas particles currently making up
that phase. Figure \ref{fig_epoch} shows the redshift dependence of
the age of enrichment: the left panel shows the enrichment epoch of
the WHIM phase in units of cosmic time at the same redshift, $\bar
t_{\rm phase}(z) / t_{\rm cosmic}(z)$, for the same simulations of
Figure \ref{fig_phasespace}.

Figure \ref{fig_epoch} shows the enrichment epoch of
the WHIM phase. We note that it decreases with time in all runs, thus
implying that at high redshifts the metal mass deposition takes place
at a larger fraction of the cosmic time. Thus, the lower the value of
$\bar t_{\rm phase}(z) / t_{\rm cosmic}(z)$, the earlier the
enrichment of the particles belonging to that phase  happened. We also point
out that the rate at which the ratio $\bar t_{\rm WHIM}(z) / t_{\rm
  cosmic}(z)$ decreases with redshift provides a measure of when the
bulk of metals are received by the particles belonging to each
phase. For the NW simulation, the WHIM enrichment is always quite
recent and is due to the combined action of star formation and
structure formation: while the former sets the time-scale at which
metals are produced and assigned to the gas surrounding star-forming
regions, the latter determines the epoch at which potential wells are
sufficiently developed for gas dynamical processes to become effective
in removing gas from such regions during the hierarchical assembly of
structures. The W and CW runs exhibit a redshift dependence that is
quite similar to that of the NW run. This is not surprising since the
wind mass--load that carries the metals to the WHIM phase is also
proportional to the star formation rate, while gas--dynamical processes
dominate at low redshifts, when the star formation rate drops at a
similar rate in all these runs. A significantly different history of
enrichment is instead found in the BH run. In this case, the sudden
episodes of energy release, occurring around $2\mincir z \mincir 4$
when the mass accretion rate onto BHs peaks, are very effective in
heating and expelling a large amount of enriched gas from the
galaxies. Indeed, the faster decrease of the $\bar t_{\rm WHIM}(z) /
t_{\rm cosmic}(z)$ ratio for the simulation with BH feedback is a
consequence of the different timing of the enrichment episodes of the
particles that then ended in the WHIM. Therefore, while winds produce
an enrichment which starts earlier and proceeds more gradually,
AGN-BH feedback heats and enriches the particles at later epochs and
over a shorter time interval. On the other hand, the lack of feedback
capable to displace enriched gas in the NW run delays the release of
metals from star--forming regions and closely links the timing of
enrichment for WHIM and warm phases to the timing of growth of large
potential wells, where stripping of enriched gas from merging haloes
takes place.

Another useful diagnostic to study the enrichment timing of the IGM is
represented by the relative delays of enrichment between the WHIM and
warm phases and between the {\it diffuse} and {\it collapsed}
environments. We recall that we defined a gas particle to belong to
the diffuse phase whenever it has density contrast $\db<50$, while it
belongs to the collapsed phase for $\db>50$. 

In the left panel of
Figure \ref{fig_epoch_diff} we plot $\bar{t}_{\rm WHIM,\db\, < 50} -
\bar{t}_{\rm warm,\db\, < 50}$, while the right panel shows
$\bar{t}_{\rm WHIM,\db\, > 50} - \bar{t}_{\rm warm,\db \, > 50}$.  As
shown in this panel, at high redshifts
the enrichment of the diffuse WHIM phase in the NW run is on average
roughly coeval to that of diffuse warm phase, while it becomes
progressively more recent at lower redshifts (by $\sim 0.2$ Gyr at
$z\sim 2$ and by $\sim 1$ Gyr at $z=0$). As already mentioned, this
evolution is purely driven by gas--dynamical processes which bring
metals from merging haloes to the diffuse medium, whenever these haloes
enter in the pressurized medium permeating large potential
wells. Since enriched gas is extracted from galaxies and groups of
galaxies at temperatures typically hotter than that of the IGM, cooler
diffuse enriched particles need more time to reach lower
temperatures. This implies that they have been stripped earlier than
hotter gas, thus explaining why warm diffuse gas has been enriched
before WHIM diffuse gas.

Quite interestingly, the presence of galactic winds (W and CW runs) is
effective in improving the mixing between the warm and WHIM diffuse
phases, thus making their enrichment coeval down to lower redshifts,
$z\simeq 1$. At lower redshifts, gas stripping within large structures
becomes again the dominant process, thus recovering the same
qualitative behaviour as in the NW run. As for the run with BH
feedback, its behaviour is quite close to that of the NW run at high
redshifts. At $z\mincir 3$ gas, displacement from already enriched
haloes to the WHIM phase becomes gradually more efficient, thus making
the enrichment of the diffuse WHIM more recent than for the diffuse
warm gas.

As for the gas in the collapsed phase ($\db>50$, right panel), the
enrichment of the warm and of the WHIM phases is almost coeval down
to $z\simeq 2$ in the NW run. The reason for this lies in the
relatively short time scale over which warm particles within haloes
are heated to $T>10^5$K and enriched when they approach star forming
regions. At lower redshift, the growth of progressively larger
structures makes gas in a shock-heated phase to have a progressively
higher temperature, until it eventually reaches the WHIM temperature
range. At $z=0$ this gas makes up a phase with $T>10^5$K, reaching
overdensity of up to $\db\sim 10^4$, as shown in the upper left panel
of Fig. \ref{fig_phasespace}. Differently from the warm dense gas, the
shock-heated WHIM does not lie close to star forming regions and, as
such, it did not experience recent enrichment episodes. By $z=0$, the
enrichment age of the WHIM takes place on average $\magcir 1$ Gyr
earlier than that of warm gas. As for the W run, we note that the
effect of winds is that of shortening the time scale that warm gas
takes to reach WHIM temperatures, thereby making even more coeval the
enrichment age of the diffuse and collapsed phases. The run with BH
feedback is again very close the NW run at $z\magcir 3$. At lower
redshift, BH accretion becomes efficient. The subsequent energy
feedback causes a fast removal of recently enriched gas from the very
dense warm phase, with $\db>10^4$, surrounding star-forming
regions. This effect is visible in Fig. \ref{fig_phasespace}: comparing
upper left and lower right panels one clearly notices the depletion of
dense and warm gas in the BH run. This gas is shock heated to larger
temperatures, thus providing a supply of recently enriched medium to
the WHIM phase.

In general, these results show that the timing of enrichment of the WHIM
does depend on the nature of the feedback included in the
simulations. The presence of winds leaves its fingerprint in the
timing of enrichment of the diffuse component, with $\db<50$, at
relatively high redshift, $z\magcir 1$. On the other hand, BH feedback
has a much more evident effect at relatively low redshift, $z\mincir
2$, i.e. in the regime where it reaches the peak of efficiency in
quenching star formation and in displacing gas from star forming
regions. Quite interestingly, the effect is more pronounced for the
collapsed gas phase, $\db>50$. This suggests that studying the
enrichment pattern of galaxy clusters and groups, out to their
outskirts, should represent the best diagnostic for the role played by
AGN feedback in determining the cosmic cycle of metals
\citep[e.g.,][]{fabjan_etal09}.

\subsection{Properties of the WHIM in the local Universe}
\label{whim_properties}
\subsubsection{Total and metal mass distribution as a function of
  density and temperature}

\begin{figure*}
\includegraphics[width=0.45\textwidth]{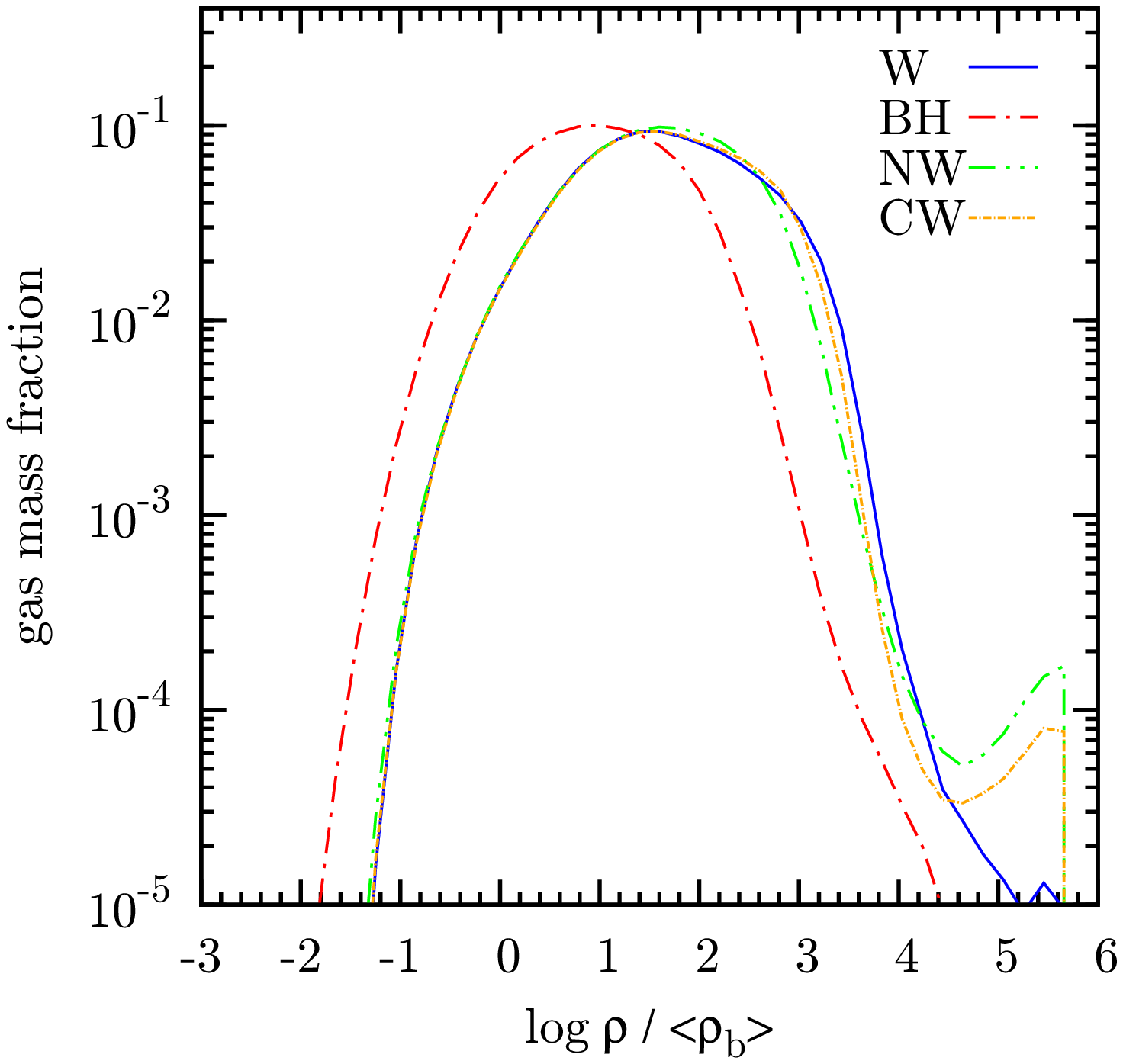}
\includegraphics[width=0.45\textwidth]{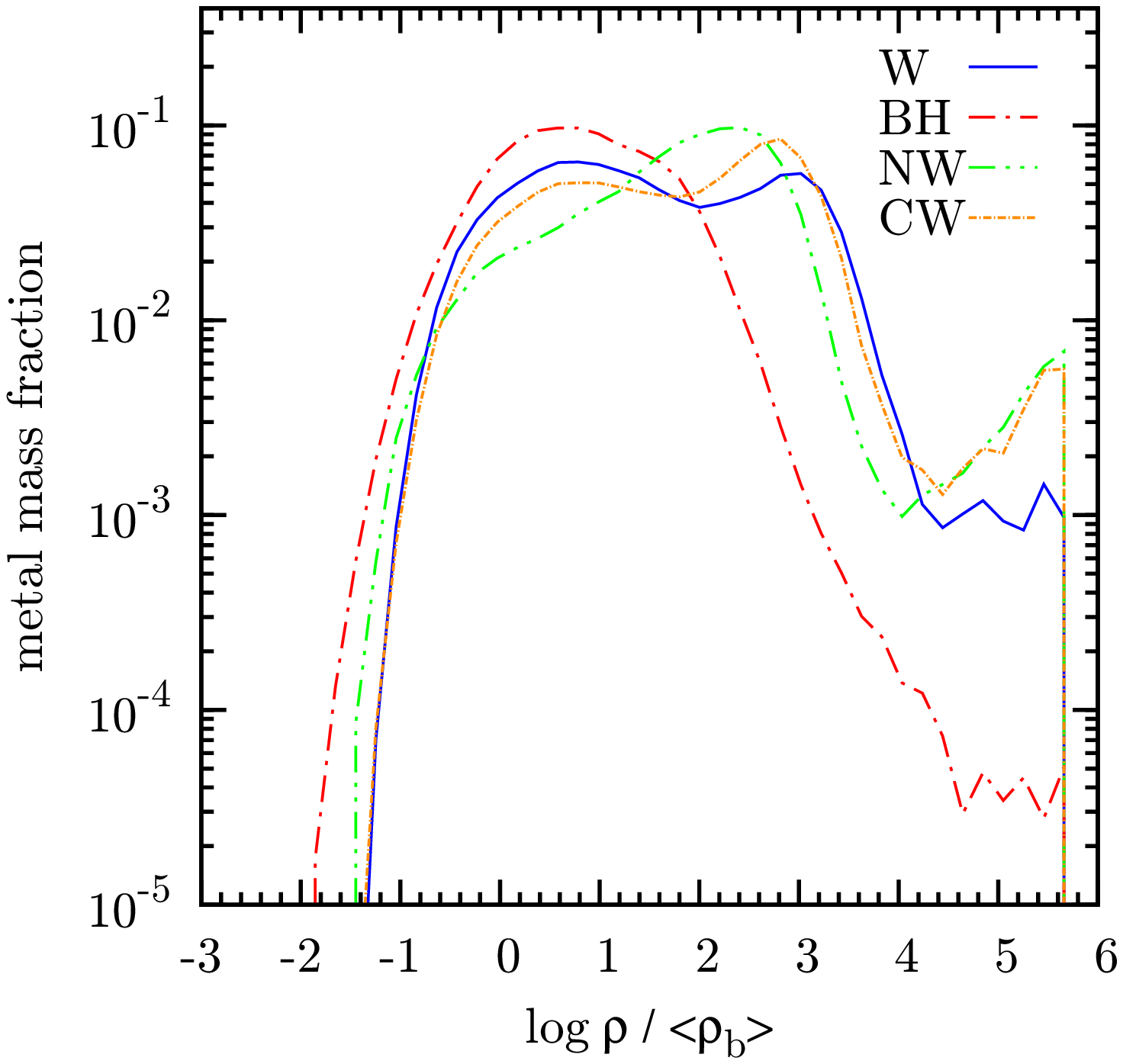}
\caption{Probability distribution functions (PDF) of the total
  WHIM mass (left panel) and WHIM mass in metals (right panel) at
  $z=0$ as a function of gas density (in units of the mean cosmic
  baryon density) for the W, NW, CW, BH simulations, represented by
  the  continuous blue, dot-dashed green, double-dot-dashed orange
  and dashed red lines, respectively.}
\label{fig_whimdensity}
\end{figure*}

\begin{figure*}
\includegraphics[width=0.45\textwidth]{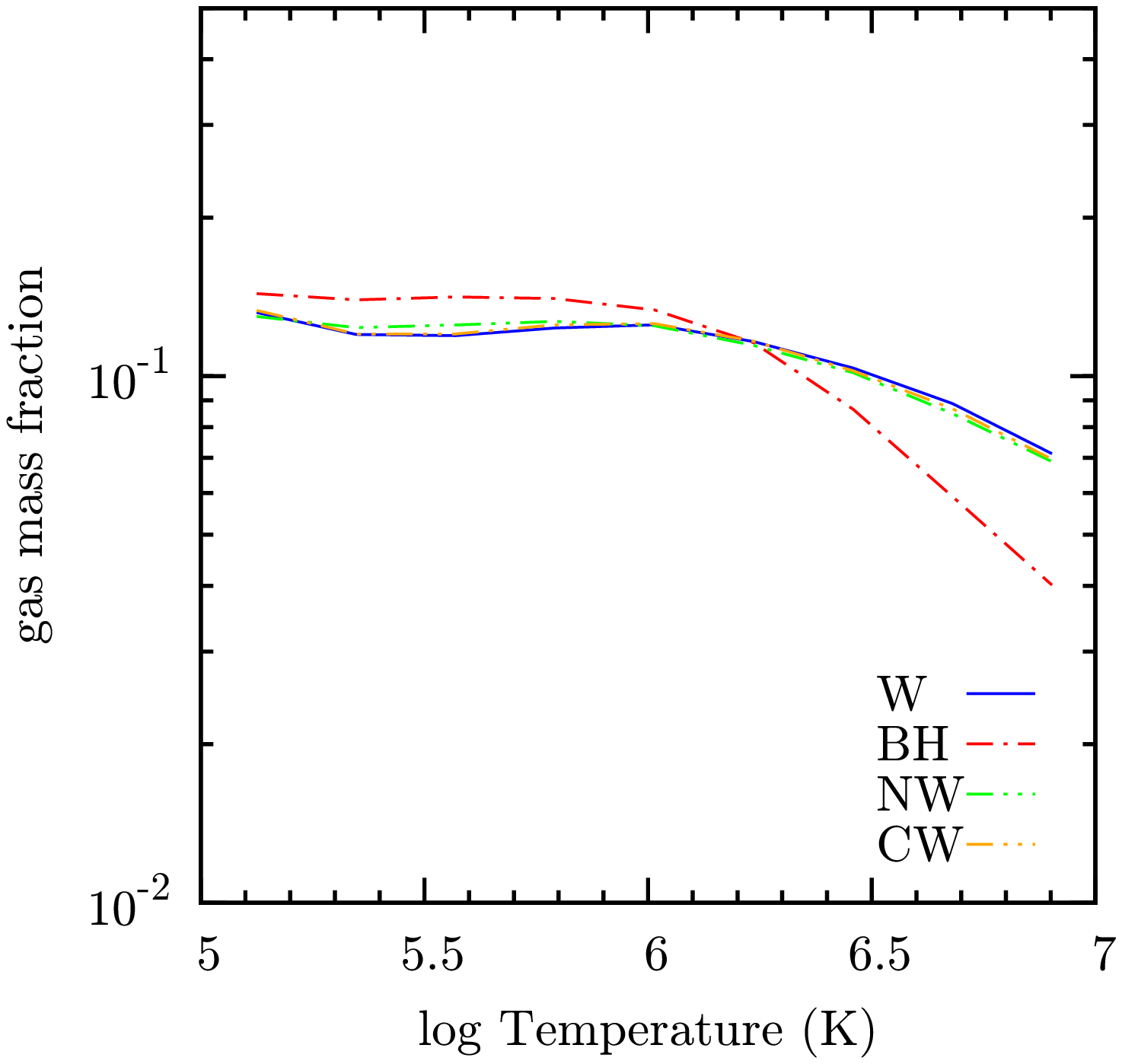}
\includegraphics[width=0.45\textwidth]{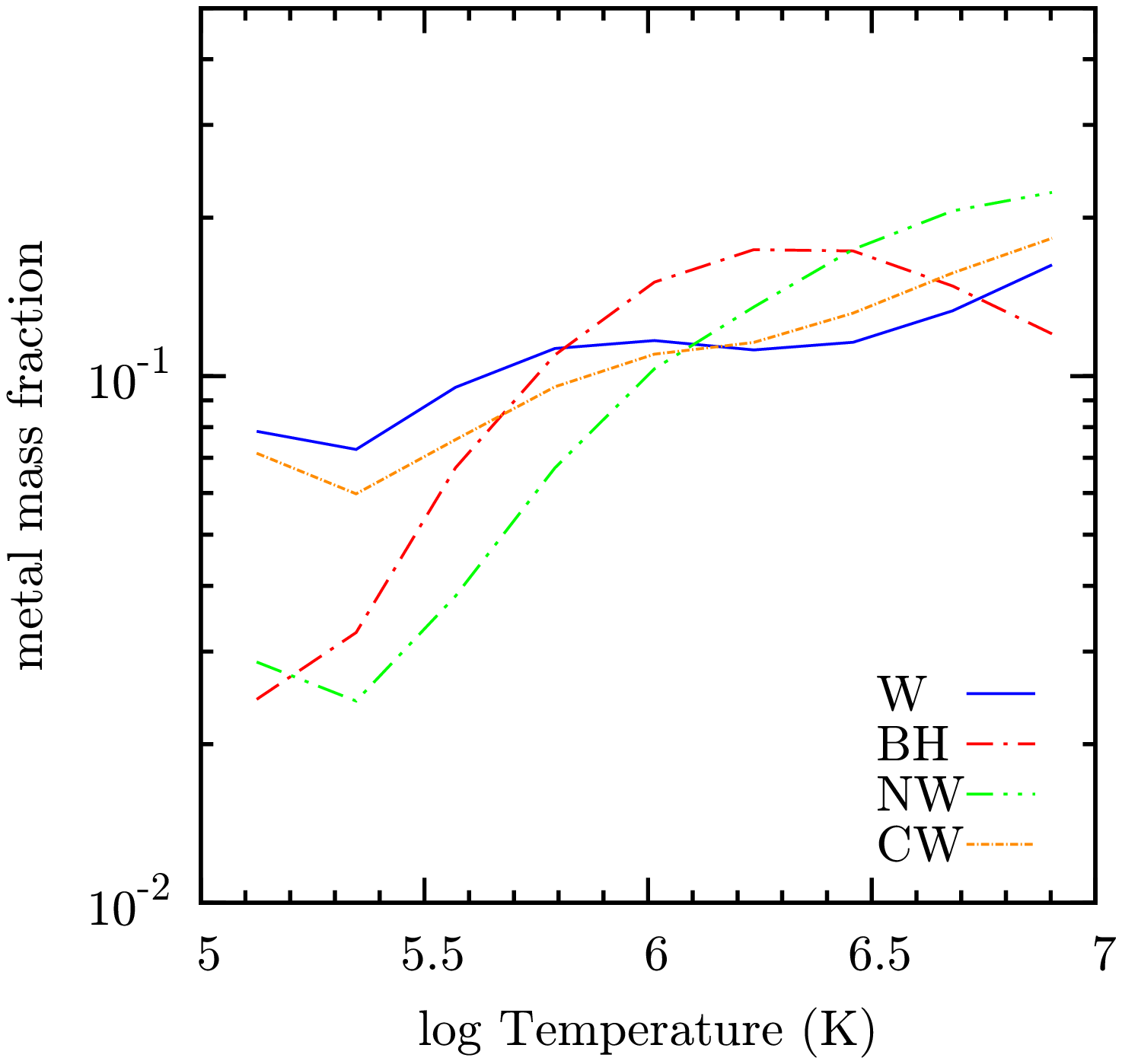}
\caption{Probability distribution functions of the rotal WHIM
  mass (left panel) and WHIM mass in metals (right panel) at $z=0$ as
  a function of gas temperature  for the W, NW, CW, BH
  simulations, represented by the  continuous blue, dot-dashed
  green, double-dot-dashed orange and dashed red lines, respectively.}
\label{fig_whimtemperature}
\end{figure*}

While in the previous sections we analysed the evolution of different
phases, defined according to temperature criteria, we present now an analysis of the metal and gas content of the
WHIM at $z=0$ as a function of gas density and temperature. As such,
this analysis has implications for the detectability of the WHIM in the
local Universe and for connecting its properties to the feedback
mechanisms which determine its density and metallicity structure.
The results are presented in Figures \ref{fig_whimdensity} and
\ref{fig_whimtemperature} for the density and temperature
distributions, respectively. In the left panels of these figures we
show the probability distribution function (PDF) for the total WHIM
mass, while in the right panels we show the corresponding PDF for the
metal mass in the WHIM. These PDFs are defined so that they provide
the fraction of WHIM gas mass (metal mass) contributed by unit
logarithmic intervals in gas density (Fig. \ref{fig_whimdensity}) and
temperature (Fig. \ref{fig_whimtemperature}).

Consistently with the phase diagrams shown in
Fig. \ref{fig_phasespace}, the left panel of Fig. \ref{fig_whimdensity}
shows that BH feedback predicts a mass distribution for the WHIM
component which is quite different from the other simulations: it
reaches smaller underdensities, while the high--density
tail is suppressed.  The reason for this is that this feedback model provides
episodes of strong heating of the gas around BHs, allowing it to reach
low density regions, thus reducing the gas content of virialised haloes. At $z=0$,
we find that $\sim 50$ per cent of the WHIM in the NW, W and CW runs has
$\db<50$, thus lying outside collapsed structures. This fraction
increases to $\magcir 80$ per cent in the run with BH
feedback. Correspondingly, the peak of the WHIM density distribution
in the BH case lies at $\db\simeq 10$, a value that increases to
$\db\simeq 80$ in the other simulations.  This result is in line with
that found by \cite{bhatta08} and \cite{puch08}, who showed that BH feedback is
effective in removing gas from virialized haloes having masses typical
of galaxy groups. Differences between the W, CW and NW runs are
instead more prominent in the high density tail of the distribution,
$\db > 10^4$, where they involve in any case only a tiny fraction of
the total WHIM mass. In the NW and CW runs, gas accumulates around star
forming regions, reaching densities close to the threshold density for
the onset of star formation.  The effect of the decoupled winds in the
W run is that of efficiently removing this gas and bringing it to
lower density regions, with $\db\simeq 10^3-10^4$.

Similar features that describe the WHIM mass distribution are also
visible in the distribution of the mass in metals (right panel of
Fig. \ref{fig_whimdensity}). Metals are more likely found at
overdensity $\db \sim 10$ in the BH simulation, while for the NW case
this overdensity is usually between $10^2$ and $10^3$. Again, this
comparison clearly shows that BH feedback is highly efficient in
displacing metal--enriched gas from high density regions, thereby
providing a more uniform enrichment of the WHIM at $z=0$. The PDFs for
the CW and W runs have a sort of double--peaked shape: one peak is at
moderate overdensity, $\db \sim 10$, that corresponds to the WHIM
lying outside virialized haloes, where winds typically deposit
metal--enriched gas; the second peak is at $\db \sim 10^3$, well inside
virialized haloes and close to star-forming regions. Such a
double--peak structure is more evident for the W run, where winds are
more efficient in escaping high--density regions.  Finally, we note
that in the BH case the metal mass fraction at $\db > 10^2$ is
dramatically smaller (by up to 2 orders of magnitude) than in all the
other cases.

As for the temperature distribution of the WHIM, shown in the left
panel of Fig. \ref{fig_whimtemperature}, we find that winds have
a negligible effect. For BH feedback, it has the effect of reducing
the mass fraction of gas above $10^6$K, while correspondingly
increasing the fraction of gas at lower temperature. Indeed, BH
feedback is efficient in removing gas from haloes having virial
temperature within the WHIM range. After being heated by BH feedback, this
gas leaves the halo potential wells and cools to lower temperature,
$T<10^5$K, by adiabatic expansion.

{ In general, the left panels of Figs. \ref{fig_whimdensity} and
  \ref{fig_whimtemperature} highlight that the presence of galactic
  winds has a negligible impact on the distribution of gas mass in
  both density and temperature. Although the presence of BHs has a
  more apparent effect on the distributions of gas mass, it is
  clear that the distributions of metal mass fractions are much more
  sensitive to the adopted feedback model (see right panels of
  Figs. \ref{fig_whimdensity} and \ref{fig_whimtemperature}).}  This
is not surprising, since feedback mostly impacts on the gas
surrounding star forming regions, which has the higher enrichment
level. In the NW run, the metal mass fraction in gas at the
low-temperature boundary of the WHIM phase, $T\magcir 10^5$K, is about
one order of magnitude lower than for the hot WHIM at $T\mincir
10^7$K. The effect of winds is that of transporting metal rich gas
from high density regions, where it reaches high temperature, to lower
density regions, where is cools down by adiabatic expansion. As
expected, hydrodynamically coupled winds (CW) make it harder for the
gas to leave the densest regions, thus producing a distribution which
is intermediate between those of the NW and W runs.

\subsubsection{The distribution of the WHIM metallicity} 
Having characterized how the mass in gas and in metals is distributed
as a function of the WHIM density and temperature, we describe now how
the WHIM metallicity depends on density and temperature. This
characterization of the WHIM is more observationally oriented. Indeed,
different distributions of metallicity determine what are the best
tracers (chemical elements and their ionization states) to reveal the
presence of the WHIM and characterize its physical properties: the
presence of different ionization species, that are thought to be the
best WHIM tracers, depends in fact upon local conditions of density,
temperature and ionization field \citep[see][for a review]{richter08}.

\begin{figure*}
\includegraphics[width=0.45\textwidth]{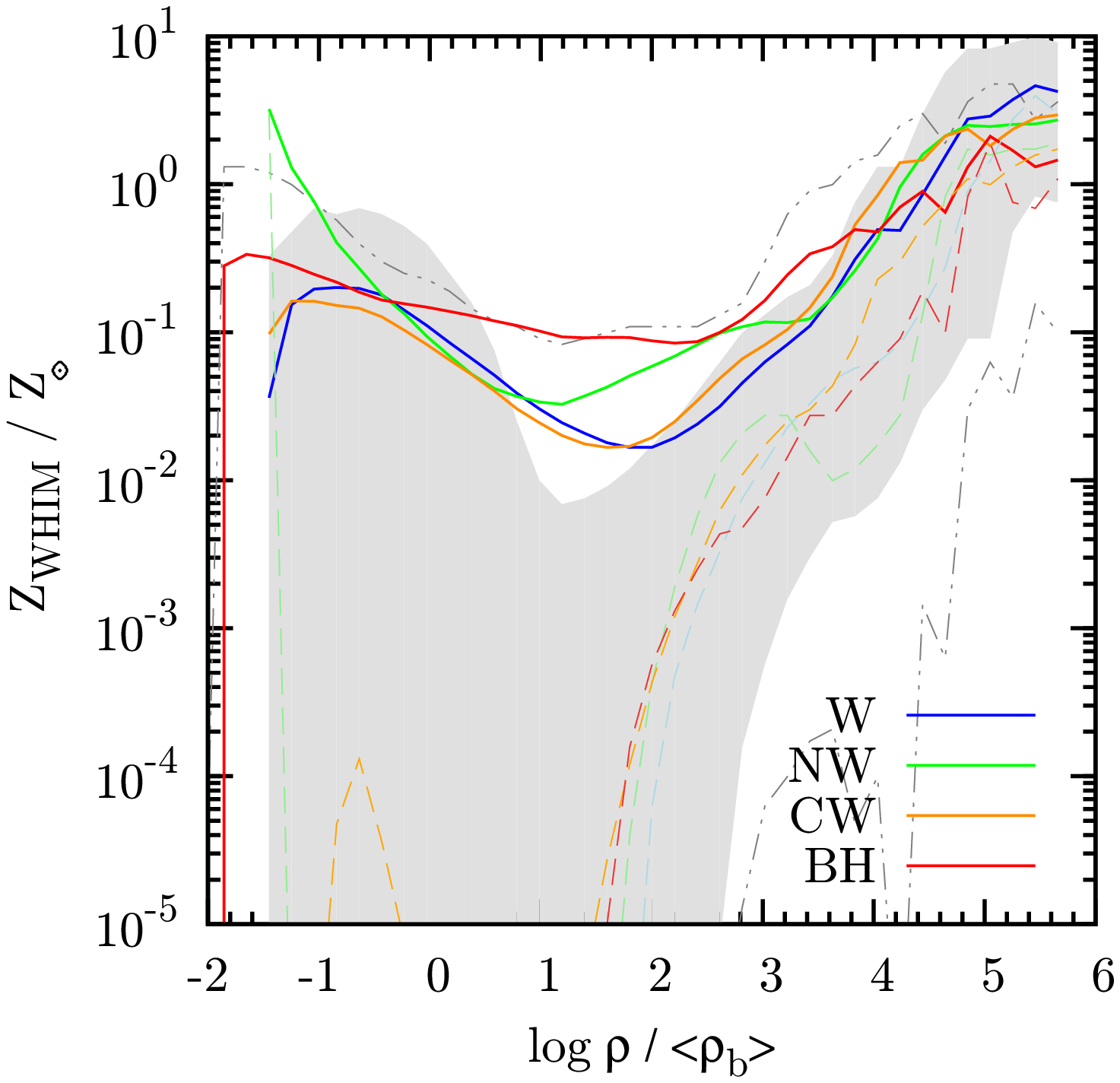}\includegraphics[width=0.45\textwidth]{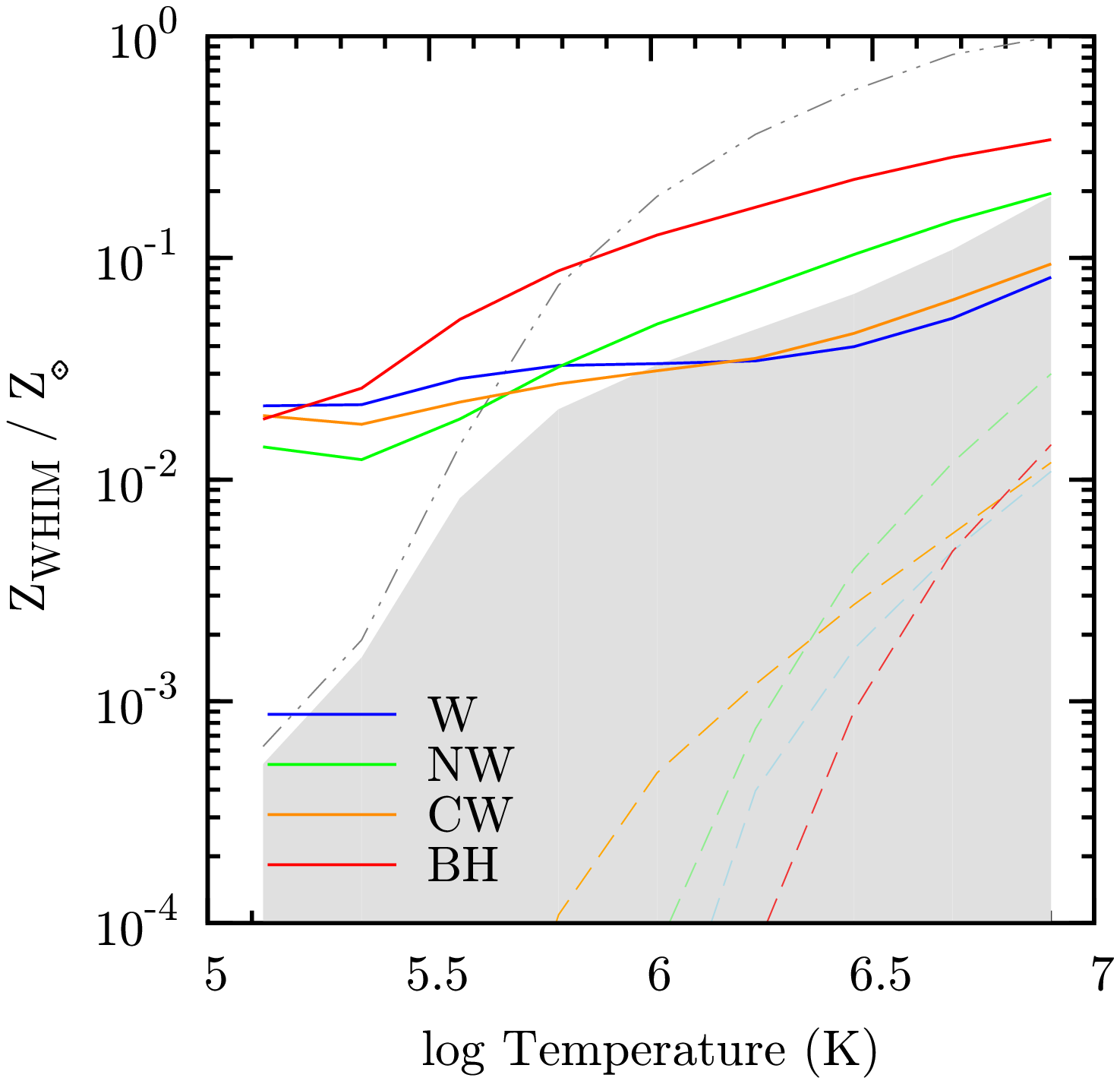}
\caption{ Total metallicity of the WHIM (gas particles at temperatures
  $10^5-10^7$ K) in solar units at $z=0$, as a function of gas density
  (in units of the cosmic mean baryon density $\lb \rho_{\rm b} \rb$,
  left panel) and temperature (right panel).  { In each panel, the
    grey shaded area encompasses the 10 and 90 percentiles of the W
    run, while the dot--dot--dashed lines show the same percentiles
    for BH the run. Thick coloured lines show the average
    metallicities while thin dashed lines show the median
    metallicities}.}
\label{fig_density_temperature_metallicity}
\end{figure*}

We show in Figure \ref{fig_density_temperature_metallicity} the
distribution of the total metallicity as a function of density within
the WHIM temperature range { (left panel)} and as a function of
temperature { (right panel)}. Since the metallicity distributions
have always a quite large scatter, a meaningful way of presenting the
results is in terms of the { mean} (continuous lines), { median
  (dashed lines)} and of the 10 and 90 percentiles { (shaded area
  for the W run, dot--dot--dashed lines for the BH run)} of these
distributions. For reasons of clarity, we do not report in these
figures the scatter for the NW and CW run.

As for the density dependence of metallicity in the W, CW and NW
simulations (upper left panel), we note that at overdensities between
10 and 100, which is the range where the PDF of the WHIM mass
distribution reaches its maximum value (see Fig.
\ref{fig_whimdensity}), the median metallicity has a strong positive
correlation with gas density and increase from $\log (Z/Z_\odot)\simeq
-6$ to $\log (Z/Z_\odot)\simeq -2$. Furthermore, the overall scatter
around these median values is about two orders of magnitude, so that
metallicity values as large as $0.1Z_{\odot}$ are not unlikely. An
obvious observational implication of this large scatter in the
metallicity distributions is that a fairly large number of
lines-of-sight are required, along which measuring WHIM metallicity,
in order to properly populate such a scatter. { It is worth
  mentioning that the large difference between the average and median
  metallicities in regions with $\db \mincir 1$ arises from the highly
  skewed distribution of metallicity in this regime: most of the
  underdense gas is not enriched at all, while metals mainly lie in a
  small number of highly enriched particles with $Z\magcir 0.01$. This
  is mostly true for BH and NW runs, while in W and CW runs the
  distribution of metals with particle metallicity is significantly
  shallower. We point out that highly enriched particles preserve
  their metal content due to the intrinsic lack of diffusivity of the
  SPH. Including an explicit description of metal diffusion
  \citep[e.g.,][]{greif09} would make such particles sharing their
  metal content with the surrounding metal--poor particles.}

As for the metallicity distribution as a function of temperature
(bottom panels), we note a clear trend for metallicity to increase
with temperature in all cases, although the details of this trend have
a dependence on the feedback model. The median metallicities in the W,
CW and NW runs (bottom left panel) increase from
$10^{-4}-10^{-3}Z_\odot$ to $\magcir 10^{-2}Z_\odot$ as the
temperature increases from $10^6$K to $10^7$K. The NW simulation
predicts a metallicity of the hottest WHIM, with $T\simeq 10^7$K,
which is higher by about a factor three than for the W and CW
runs. This gas is located within the virialised regions of galaxy
group haloes, which have typical overdensities of about $10^2-10^3$
(see upper left panel). In the NW run, this gas, which remains at
these overdensities for a relatively long time before being stripped,
has been more heavily enriched by the more (in fact, exceedingly)
intense star formation. Quite interestingly, the CW simulation shows
instead larger median metallicity at temperatures $10^{5.5}-10^6$ K,
with a shallower decline towards lower temperatures. Gas at this
temperature is a mixture of highly enriched gas in virialized haloes
and of poorly enriched gas at low density.  As can be inferred from
the phase diagrams shown in Fig. \ref{fig_phasespace}, gas at $z=0$ in
the CW run lies within the WHIM temperature range either for
$10\mincir \db\, \mincir 100$ or right outside of the star--forming
regions, i.e.  $10^4 \mincir \db\, \mincir 10^5$, where enriched wind
particles loose their kinetic energy due to hydrodynamical
interactions.  As a consequence, in the CW run the WHIM at these
overdensities is more enriched than in other runs.

As for the comparison between NW and BH runs (right panels), we note
that BH feedback removes a lot of metal enriched gas from the
surroundings of star forming regions (see also
Fig. \ref{fig_phasespace}). This explains the lower metallicity of the
densest WHIM in the BH simulation (upper left panel). Furthermore, the
suppression of star formation in the BH run also reduces the
enrichment level of the shocked gas within collapsed regions with
$10^2\mincir \db\mincir 10^3$. At the same time, { BH feedback is
  quite effective in increasing the metallicity of gas lying outside
  virialized structures, down to rather low densities, $\db \mincir
  10$. The reason for this widespread enrichment is the high
  efficiency of BH feedback to displace highly enriched gas from the
  haloes of large galaxies in the redshift range $z\simeq 2$--4, in
  correspondence of the peak of the BH accretion rate. }

As for the dependence of the WHIM metallicity on
temperature, the NW and BH runs provides similar trends, with a lower
level on enrichment in the BH run.

In general, the results of our simulations confirm that different
feedback models leave distinct signatures in the metallicity structure
of the WHIM. While these results have in principle interesting
implications for the observational characterization of WHIM, it must
be emphasized that the large scatter expected in the metallicity
distributions requires that metallicity measurements should be carried
out for a fairly large number of lines--of--sight. We defer to a
future paper the presentation of an observationally--oriented analysis
of our simulations, where we discuss the efficiency with which the
metallicity structure of the WHIM can be recovered from mock
absorption spectra of background sources.

\subsection{Evolution of the relative metal abundances}
While the above analysis was aimed at describing the overall
metallicity of the different gas phases, a much richer amount of
information is provided by our chemo-dynamical simulations. Indeed,
the possibility of tracing the enrichment of different elements allows
us to study differences in the timing of enrichment of different heavy
elements. In this section we explore the redshift evolution of the
ratios between the abundances of Oxygen with respect to Iron and to
Carbon, with the purpose of quantifying the
signatures that IMF and feedback mechanisms leave on the
WHIM enrichment history. Atomic transitions associated to these three
elements are often used to reveal the presence of the diffuse
IGM and to measure its enrichment level at different redshifts. For
instance, CV, OVI and OVII lines are considered as the most prominent
absorption features in the far-UV and soft X-ray spectra of background
sources to reveal the presence of the WHIM in the nearby Universe
\citep[e.g.,][for a review]{richter08}. At the same time, CIII, CIV
and OVI absorption lines are commonly used to trace the metal content
of the IGM at $z\magcir 2$
\citep[e.g.,][]{SchayeCarbon03,pieri04}. Finally emission features
associated to the OVIII line and to the Fe-L and Fe-K complexes are
used to trace the metal content of the intra-cluster and intra-group
media \citep[e.g.,][ for a review]{werner08}.

Since different elements are produced in different proportions by
different stellar populations, the evolution of their relative
abundances is expected to depend, for a fixed mass--dependent
life-time function, on the shape of the stellar IMF (e.g.,
\citealt{T07,wiersma_etal09}; see \citealt{Borgani08} for a review).
In general, we expect that different spatial distributions
characterize the enrichment pattern for different elements. As an
example, products of SN-II are released over a shorter time-scale than
those arising from SN-Ia. Since star particles are expected to move
from their original location where they have formed, e.g. due to
merging or stripping processes, we expect that SN-II products pollute
gas particles lying very close to star forming regions, while SN-Ia
products should have a relatively more diffuse distribution. { We
  expect this effect to be more apparent in the dense environment of
  galaxy clusters where a population of diffuse inter-galactic stars
  is generated by dynamical processes. Indeed, \cite{T07} found from
  chemo-dynamical simulations of galaxy clusters that the distribution
  within the intra-cluster medium (ICM) of the metals produced by
  SN-II is more clumpy than that provided by SN-Ia, a prediction that
  has been also confirmed by observational data
  \citep{sivan09}.} 
\begin{figure*}
\includegraphics[width=0.45\textwidth]{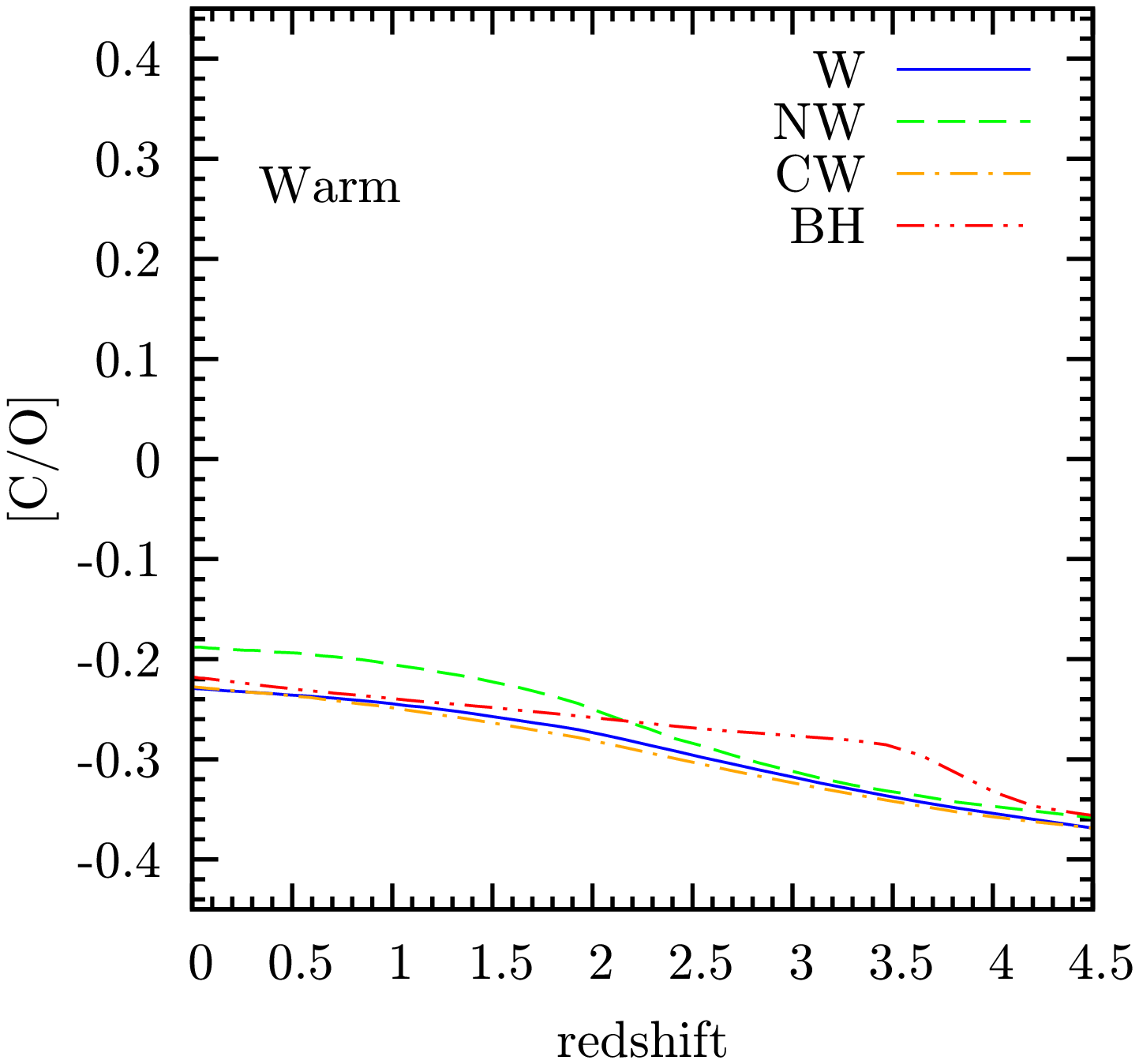}\includegraphics[width=0.45\textwidth]{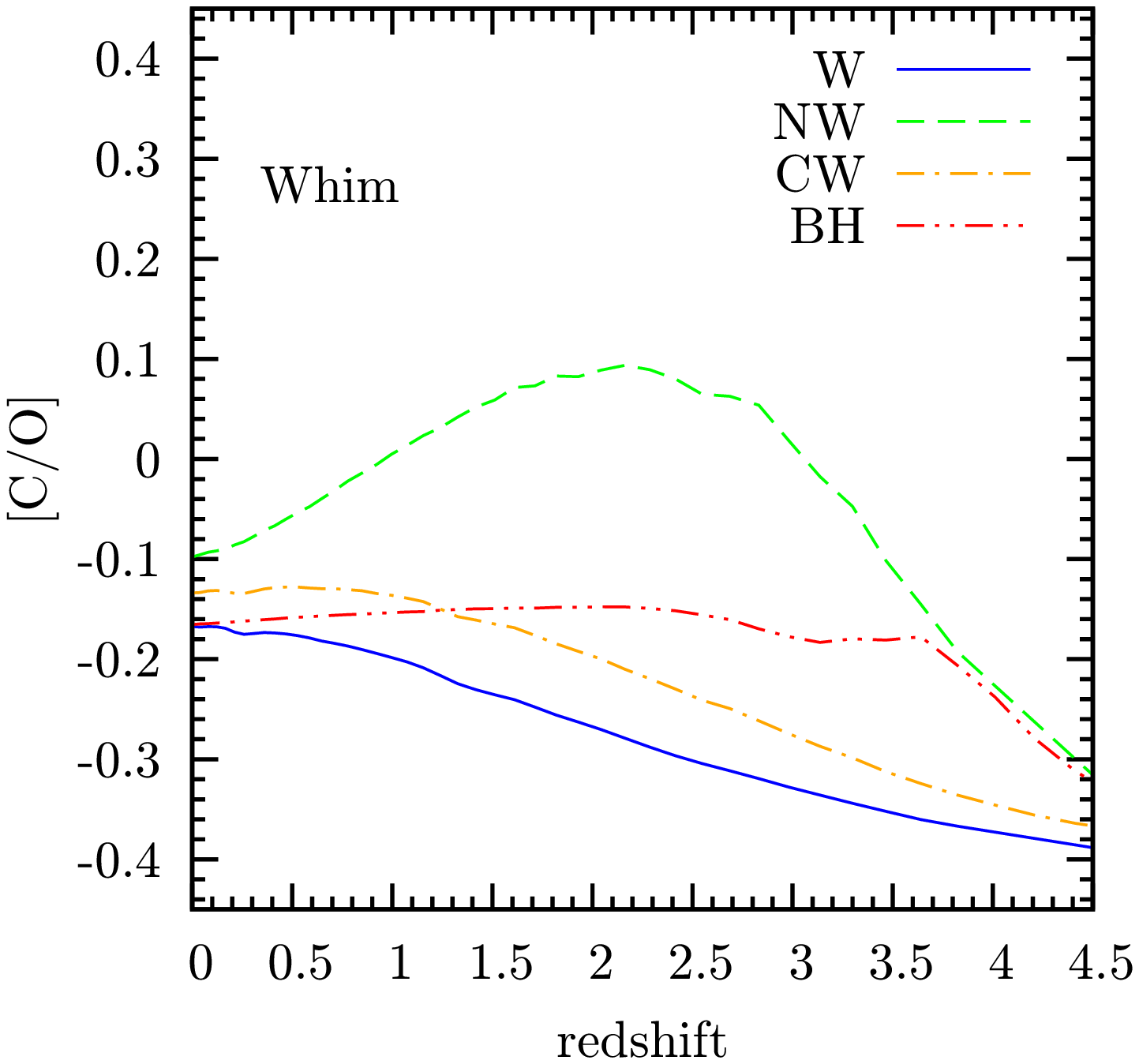}
\includegraphics[width=0.45\textwidth]{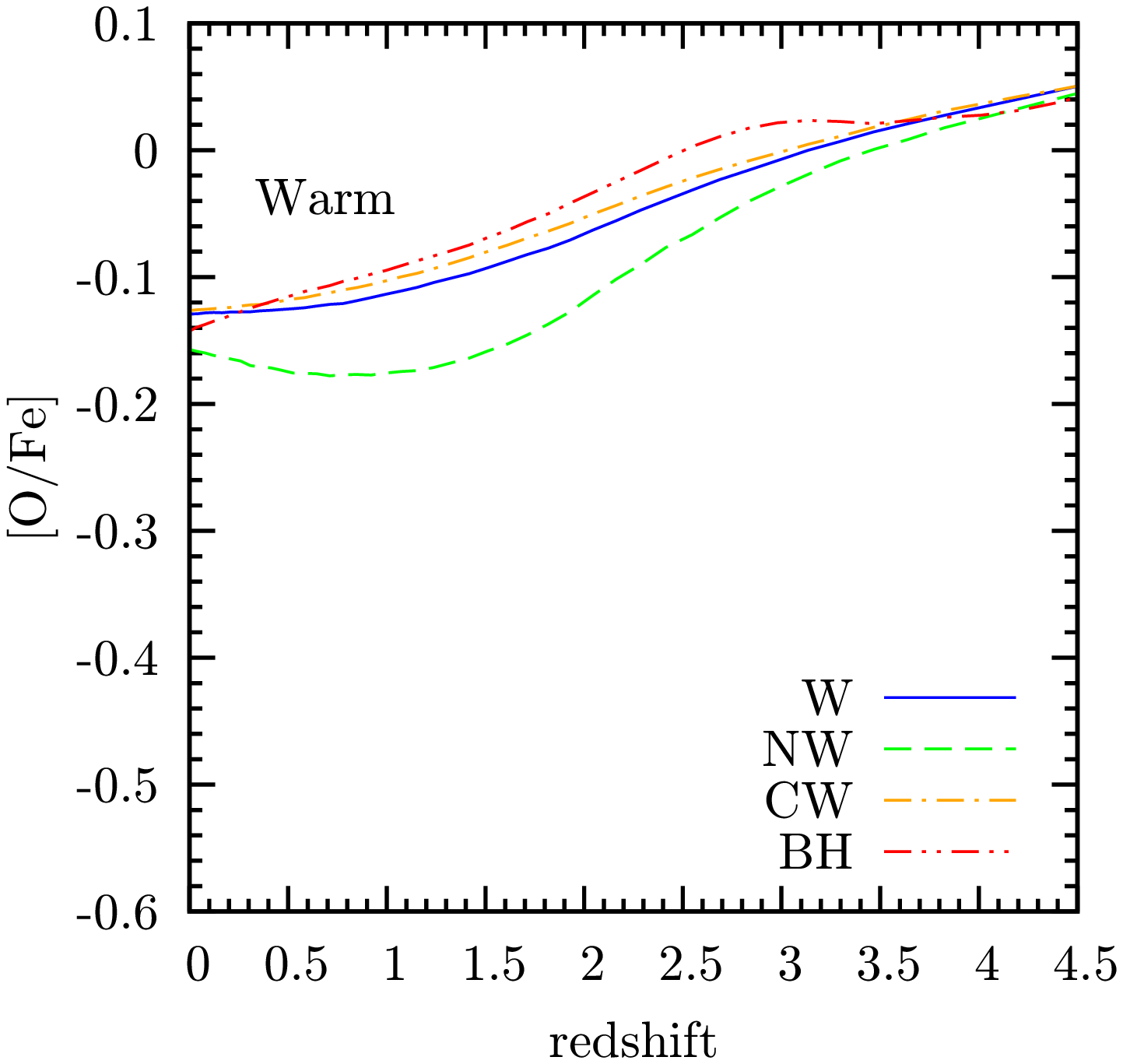}\includegraphics[width=0.45\textwidth]{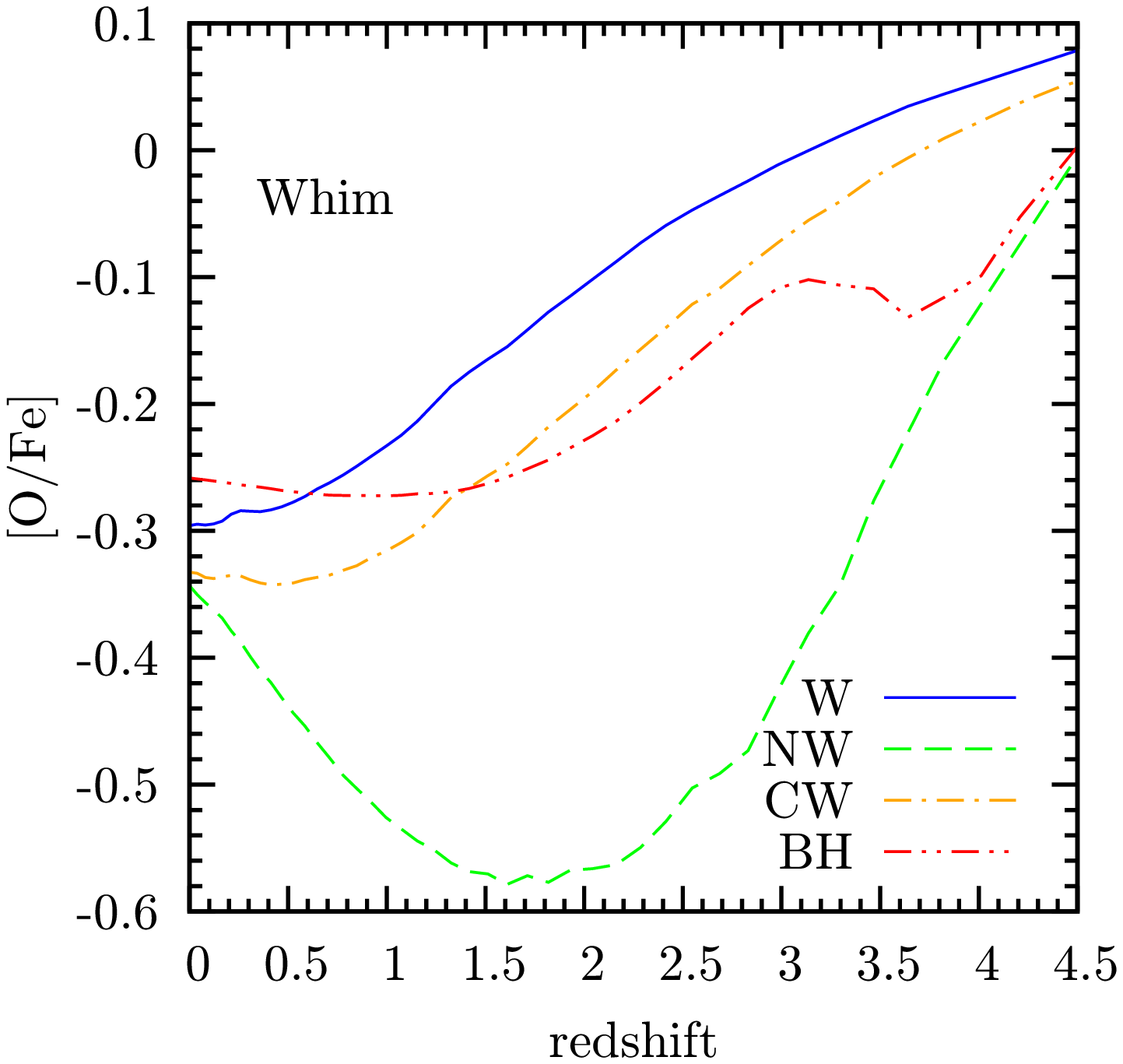}
\caption{Evolution of the mean value of the [C/O] (upper panels) and
  [O/Fe] (lower panels) ratios for the warm (left panels) and WHIM
  (right panels) phases. Each panel shows the results for the
  W$_{256,37}$, NW$_{256,37}$, CW$_{256,37}$ and BH$_{256,37}$
  simulations, represented by the continuous blue, dashed green,
  dot-dashed orange and double-dot dashed red lines, respectively.}
\label{fig_elements}
\end{figure*}

In Figure \ref{fig_elements} we plot the [C/O] and [O/Fe]
ratios\footnote{Following a standard notation, we define the relative
  abundance between the elements $X$ and $Y$ as
  $[X/Y]=\log(Z_X/Z_Y)-log(Z_{X,\odot}/Z_{Y,\odot})$.} as a function
of redshifts for the four models and for both the warm (left panels)
and WHIM (right panels) phases. These three elements are produced in
different proportions by SN-II and SN-Ia: Oxygen is almost entirely
produced by SN-II, and Iron is largely provided by SN-Ia; while Carbon
is produced in comparable proportions by SN-II and SN-Ia, the main
contributors to it at timescale comparable to those of SN-Ia are AGB
stars. The [C/O] ratio has been widely investigated in a variety of
environments, from dwarf galaxies \citep{garnett95}, to low redshift
IGM as probed by QSO absorption lines \citep{danforth08}, to high
redshift IGM at $z=2-4$ \citep{AguirreOxygen08}. As for the [O/Fe]
ratio, it is traced in emission with X-ray spectroscopy of relatively
poor galaxy clusters and groups \citep[e.g.,][]{rasmussen_ponman07}
and is used to study the relative role played by SN-II and SN-Ia in
enriching the ICM.

\begin{figure*}
\includegraphics[width=0.45\textwidth]{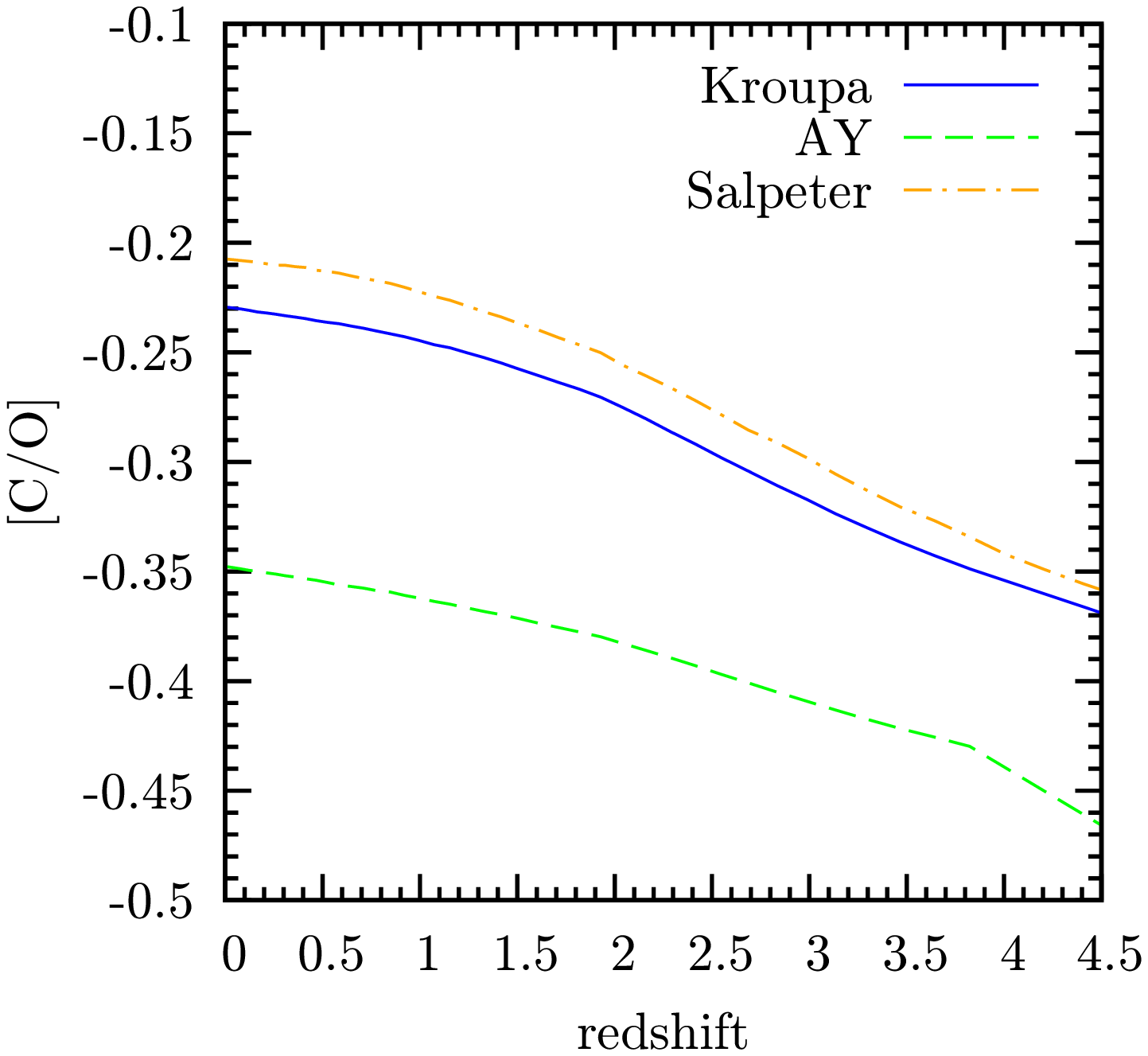}\includegraphics[width=0.45\textwidth]{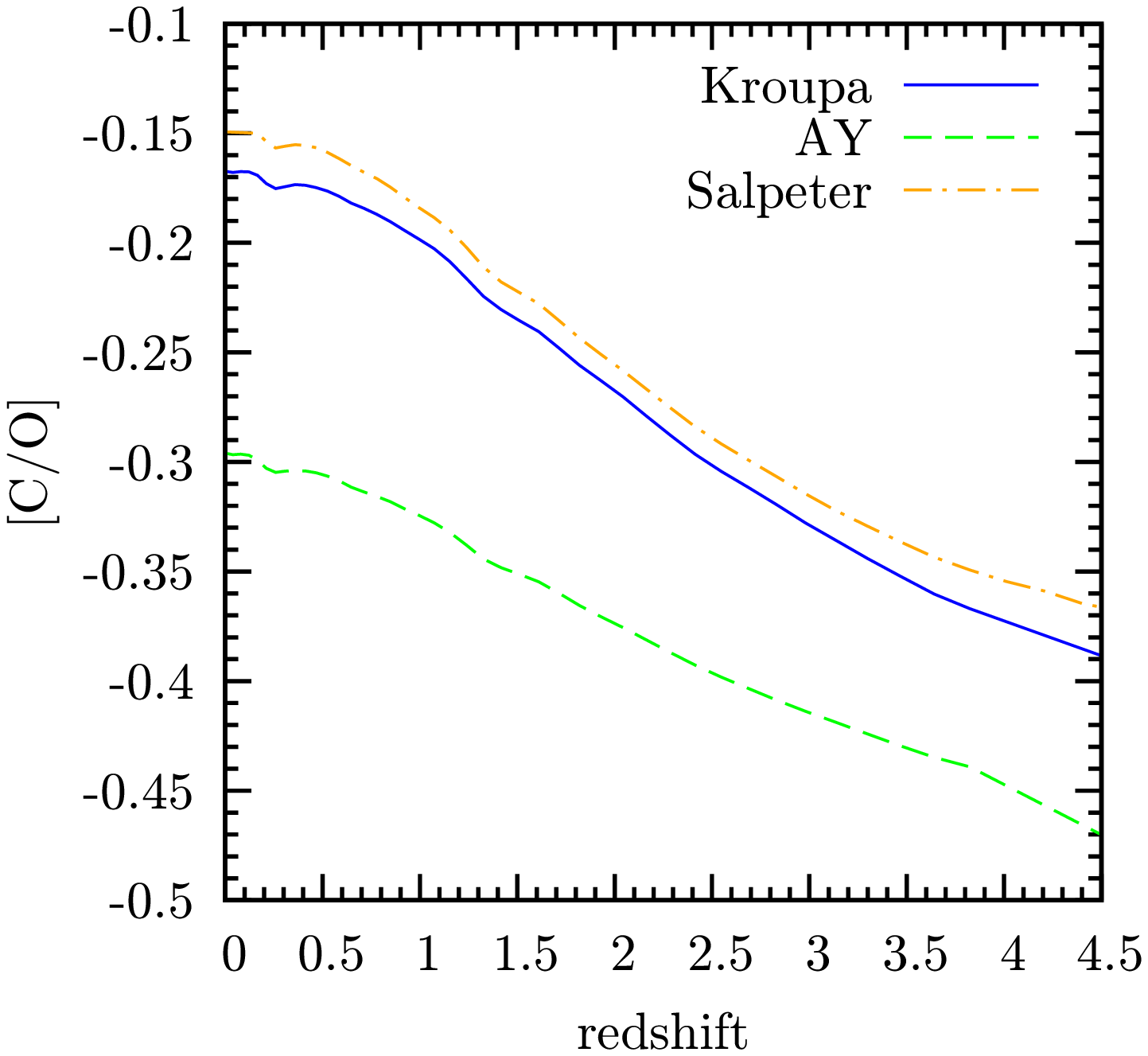}
\caption{The effect of changing the IMF on the evolution of the
  $[C/O]$ ratio for the warm (left panel) and WHIM (right panel)
  phases.  The solid blue curves are for the reference run based on
  the IMF by \protect\cite{Kroupa01} (W$_{37,256}$ run), the dashed
  green curves are for the top--heavy IMF by \protect\cite{ay87}
  (Way$_{37,256}$ run) and the dot-dashed orange curves are for the IMF by
  \protect\cite{salpeter55} (Ws$_{37,256}$ run).}
\label{fig_IMFs}
\end{figure*}

Despite the fact that different feedback prescriptions induce rather different
evolutions for the [C/O] ratio, especially for the WHIM phase, the
values attained at $z=0$ are rather similar for all the
simulations. We find [C/O]$\simeq -0.2$ and [C/O]$\simeq -0.15$ for
the warm and  for the WHIM phases, respectively. These values can
be compared with a similar analysis of chemo-dynamical simulations
performed by \cite{oppe08}. At $z=0$ they found [C/O]$\simeq -0.09$
and [C/O]$\simeq -0.05$ for the warm and warm--hot phases,
respectively, thus about 0.1 dex higher than our results (note that we
rescaled their Anders \& Grevesse 1989 abundances to our used
values). A possible reason for this difference is that \cite{oppe08}
used a model for momentum--driven galactic ejecta. However, since they
obtain rather stable results for a variety of feedback schemes, it is
not clear whether this is a likely explanation for the
difference. Another possibility lies in the different sets of yields,
mass limit for the SN-II progenitors ($10$M$_\odot$, instead of
$8$M$_\odot$ as in our simulations), and the mass--dependent
life--time function adopted by \cite{oppe08}. For instance, assuming a
larger limiting mass for SN-II turns into a relatively lower amount of
Oxygen produced, which could explain the larger value of [C/O] found
by \cite{oppe08}.

As for the evolution of the WHIM phase in the BH run, we note that
[C/O] is very close to the NW values at high redshift. This is
expected, since at early times BH accretion is quite ineffective. At
these high redshifts, gas at WHIM temperatures starts involving
shock--heated gas in filaments. This diffuse gas has been relatively
more enriched by long--lived stars that had time to move away from the
highest density star-forming regions. This explains why the relative
abundance of Carbon increases with respect to Oxygen, a trend that
extends in the NW run down to $z\simeq 2$, when a nearly solar
relative abundance is attained. After this redshift, the diffuse gas
whose temperature reaches the WHIM values, has a progressively lower
degree of enrichment. At the same time, the large amount of
low-redshift star formation in the absence of feedback causes a prompt
release of Oxygen from short-lived stars, thus motivating the gentle
decline of the [C/O] ratio. As for the run with BH feedback, the WHIM
starts receiving at $z\simeq 3.5$ a significant contribution from gas
expelled from haloes where gas accretion onto BHs reaches the peak of its
efficiency (see also the central panel of Fig.
\ref{fig_massfraction}). At lower redshift the star formation is
strongly quenched in the BH run, thus explaining the substantial
flattening of [C/O] for both the warm and the WHIM phases.

The evolution of the [C/O] ratio for the CW and W simulations is
similar for the warm and the WHIM phases, for which it rises gently by
0.1 dex and 0.2 dex, respectively, from $z=4$ to $z=0$. Differently
from the BH feedback, winds start affecting the level
of star formation already at high redshift, $z>4$. This explains the
slower increase of [C/O], which extends down to low redshift, for both
the warm and the WHIM phases. While the W and CW runs provide very
similar results for the warm phase, the tendency of the decoupled
winds to transport more efficiently enriched gas outside haloes
justifies the slightly larger values of [C/O] in the WHIM phase, as
found in the CW simulation.

The values obtained at $z=2-4$ can be compared with those inferred for
the IGM by \cite{AguirreOxygen08} using the pixel-optical depth
technique in high resolution QSO spectra. Their findings of course
depend on the strength and shape of the assumed UV background. 
For a background which is made by galaxies and quasars they find
[C/O]$=-0.7\pm 0.2$ (this value refers to Anders \& Grevesse 1989
abundances and the reported errorbar corresponds to systematic
uncertainties), which is in reasonable agreement with the results of
the warm phase (left panel) that better samples the relatively low
density IGM.  From the analysis made by \cite{danforth08}, using
absorption lines in the low redshift IGM, values of [C/O] in the range
$[-1,0]$ can be inferred, depending on which ion is used as a tracer
of the metallicity. This result is again in broad agreement with the
results of our simulations.

Since Oxygen is mostly contributed by SN-II, the behaviour of the
[O/Fe] ratio has the opposite trend compared to [C/O], with an even
more pronounced redshift dependence. The decreasing trend at low
redshift is due to the fact that Fe is primarily produced by SN-Ia
that have long-lived progenitors. The values of [O/Fe] at $z=0$ are
$\sim -0.15$ for the warm phase and $\sim -0.3$ for the WHIM. At $z>3$,
this abundance ratio approaches the solar value. Our findings are in
good agreement with the [O/Fe] ratios of gas particles in the ICM of
isolated clusters of \cite{T07}, in particular values of $-0.2$ are
reached in the outer parts of galaxy clusters and this is close to the
value of the WHIM [O/Fe] ratio.  As for the comparison with the
results by \cite{oppe08}, we note that our predicted values of [O/Fe]
are about 0.4 dex smaller than their WHIM and warm phases once we
consider the difference abundances used. These differences highlight
the need to carry out detailed comparison between chemo-dynamical
codes to distinguish the effect of the different implementation of the
chemical evolution model from the effect of the different feedback
models introduced and of the prescriptions to distribute metals around
star forming regions.

As a final analysis, we verified the effect of changing the IMF on the
resulting evolution of [C/O]. Since the slope of the IMF determines
the frequency of different SN types, we expect that a top-heavier IMF
provides a relatively larger number of SN-II and, therefore, a
correspondingly lower value of [C/O].  This is indeed confirmed by the
results in Figure \ref{fig_IMFs}, which show a lower value of C/O. A
difference of about 0.15 dex is found for the top--heavy IMF by
\cite{ay87}, for both the warm and the WHIM phase, while no sizable
differences exist between the \cite{Kroupa01} and \cite{salpeter55}
IMFs. It is interesting to note, while using a top--heavier IMF
changes the value of [C/O], its redshift dependence remains
essentially unchanged. As such, the effect of changing the IMF is
different from the effect of changing the feedback model, which
instead does change the evolutionary pattern of the relative
abundances.

The main conclusion of this section is that, in spite of the dramatic
differences in the feedback schemes and star formation histories, the
C/O and O/Fe ratios conspire to reach comparable values at $z=0$ for
all the simulations, with variations of about 20--25 per cent. Larger
differences, of up to a factor $\simeq 2.5$, are instead found at
higher redshift, $z= 2.5$. Much smaller differences in the evolution
of the relative abundances are instead found by changing the IMF. This
implies that an observational determination of the [C/O] evolution
would provide information about the nature of the feedback mechanism
responsible for regulating star formation and distributing metals.

\section{Conclusions}
\label{conclusions}
We have presented results from an extended set of chemo-dynamical
simulations which have been carried out with the GADGET-2 code
\citep{springel}, including the implementation of chemical evolution
described by \cite{T07}. The analysis presented here was focussed
primarily on the low-redshift properties of the inter-galactic medium
(IGM) and its evolutionary properties, by considering both gas in a
warm phase (gas particles with temperature { in the range
  $10^4$--$10^5$ K}) and in the so-called WHIM phase (gas particles
  with temperature in the range $10^5-10^7$ K). The main purpose of
  the analysis was to quantify the effect that different feedback
  mechanisms leave in the pattern of metal enrichment. Besides a
  simulation not including any efficient feedback (NW run), we
  performed simulations including energy feedback resulting from
  accretion onto super-massive black holes (BH run;
  \citealt{springel05bh}; \citealt{dimatteo05}), and from galactic
  winds powered by supernova (SN) explosions. We considered two
  implementations of the latter, based either on locally decoupling
  winds from hydrodynamics, so as to allow them to leave star forming
  regions (W runs), or on keeping gas in the winds always
  hydrodynamically coupled with the surrounding medium (CW runs).

The main results of our analysis can be summarized as follows.
\begin{itemize}
\item[(a)] The temperature--density and metallicity--density phase
  diagrams at $z=0$ are affected by the different feedback
  prescriptions. Including galactic winds has the effect of enriching
  with metals the low--density IGM. BH feedback is even more effective
  than winds in transporting metal-enriched gas from the star-forming
  regions to the WHIM phase. BH feedback also
  causes the presence of a non-negligible amount of relatively
  low--density, metal-enriched hot gas ($10^6-10^7$ K) which is not
  present in the simulations with galactic winds.
\item[(b)] The fraction of baryonic mass associated to warm gas, with
  { $T=10^4$--$10^5$K}, which should be associated to
  UV/Lyman-$\alpha$ absorption systems, varies from $\magcir 90$ per
  cent at high redshift, $z>3.5$, to about 30 per cent in the local
  universe, with a weak dependence on the adopted feedback model. This
  is in agreement with recent observational estimates from UV
  spectroscopy \citep[e.g.][]{danforth08}. The hot phase, with
  $T>10^7$K, sums up to 3 per cent at $z=0$, again almost independent
  of the feedback model. The fraction of baryons in stars is instead
  the most sensitive to feedback: it ranges from about 10 per cent in
  the run with no feedback (NW), to about 2 per cent in run with BH
  feedback.
\item[(c)] The WHIM phase comprises about 35 per cent of the baryon
  budget at $z=0$ for the NW run, a value that increases to about 50
  per cent for the run with BH feedback. This confirms that the share
  of cosmic baryons in the different phases does depend on the assumed
  feedback model. The redshift evolution of the mass fraction in the WHIM
  also differs in the different runs, with a stronger evolution in the
  run with BH feedback.
\item[(d)] The average age of enrichment of the warm and of the WHIM
  phases differs in the different runs, by an amount which depends on
  the gas overdensity $\db$. Diffuse gas (\db $<50$) in the warm phase
  at $z=0$ is typically enriched 0.5-1 Gyr earlier than in the WHIM
  phase. As for the gas within ``collapsed'' regions, with \db $>50$,
  the warm phase is enriched 1 Gyr after the WHIM phase in all runs
  not including BH feedback. In the BH run, enrichment of the densest
  WHIM as measured at $z=0$ takes place more than $1.5$ Gyr after that
  of the warm medium. { BHs enrich the WHIM more promptly than W in the
    redshift range $2\mincir z \mincir 4$. On the other hand, winds
    are more effective at higher redshift, when BH accretion is still
    inefficient, and below $z\sim 2$, when star formation in the BH
    run has been quenched. At all epochs, the hydrodynamically coupled
    winds (CW run) provide a slightly more recent enrichment of both
    warm and WHIM phases when compared to hydro-decoupled winds (W
    run). } BH feedback provides a faster enrichment at $z>2$, while
  below this redshift winds provide a more prompt WHIM enrichment. In
  particular, the model with hydrodynamically coupled winds (CW)
  provides at $z<3$ the most prompt enrichment of both warm and WHIM
  phases.
\item[(e)] In order to address the multi-phase nature of the WHIM, we
  compute the distribution of gas and metals in the WHIM phase as a
  function of overdensity. { As a result of the strong heating
    provided by BH feedback at $z\simeq 2$--4, the} characteristic
  density of the WHIM in the BH run is a factor of a few lower than in
  the other simulations and, correspondingly, a smaller amount of gas
  is present in the dense WHIM. The same trend is also visible for the
  amount of metals present in the WHIM. Typically, most of the metals
  lie at over--densities between a few and 10 in the run with BH
  feedback, while the typical density of the metal-enriched gas is
  about two orders of magnitude higher in the run with galactic
  winds. { The reason for this is that most of the metals are
    ejected by BH heating from galaxies between $z=4$ and $z=2$. The
    high entropy level reached by enriched gas heated by BH feedback
    makes hard for it to be re-accreted within
    collapsed haloes at lower redshift. In turn, the drop of star
    formation in the BH run causes a comparable suppression
    of metal production at $z<2$}. 
\item[(f)] Underdense WHIM regions (voids) have a very low median
  metallicity, however the metallicity of individual gas particles can
  reach values { larger than $0.1Z_\odot$ within the 90
    percentile. These enriched particles have been trasported to
    underdense regions both by galactic ejecta and by the action of
    gas-dynamical processes. Due to the intrinsic lack of diffusion in
    SPH, such particles spuriously preserve their high-metal content,
    rather than diffusing metals to surrounding metal poor particles.}
\item[(g)] The values of the [C/O] and [O/Fe] relative abundances
  at $z=0$ are similar for both the WHIM and warm phases, also
  irrespective of the feedback model. Their evolutionary pattern is
  instead sensitive to the adopted feedback model. Using a
  top--heavier IMF decreases the value of the [C/O] ratio by about
  0.15 dex at all redshifts.
\end{itemize} 

{ The simulations presented in this paper have been analysed to
  study separately the effects that SN-triggered winds and BH energy
  feedback have on the evolution of the IGM. Clearly, in a realistic
  situation one expects SN and AGN feedback to be both at work and to
  cooperate in determining the cosmic cycle of baryons. However, we
  point out that our analysis was not aimed at establishing a best-fit
  feedback model, which is able to reproduce a variety of
  observational results. We aimed insted at determining the imprints
  that different feedback models leave of the properties of the IGM
  and their possible observational signatures. Furthermore, it is
  worth reminding that the parameters that we adopted for winds and BH
  feedback have been fixed by requiring each of these two feedback
  sources to reproduce a specific observational constraint: the cosmic
  star formation rate for winds \citep{springel2003} and the
  $M_{BH}$--$\sigma$ relation for BH feedback \cite{dimatteo05}. Once
  the two mechanisms are allowed to be both present in a simulation,
  they are expected to have non--trivial interplays, which necessarely
  require a re-calibration of their characteristic parameters. }

The results obtained from our analysis have interesting implications for
the detectability of the WHIM and for the possibility of
characterising its thermal and chemical properties
\citep[e.g.,][]{stocke07}. Indeed, the possibility of detecting the WHIM
through absorption lines in the spectra of background sources does not
only depend on the amount of mass in this phase, but also on how this
mass is enriched and distributed in density and temperature. With our
analysis we have demonstrated that such distributions are rather
sensitive to the adopted scheme of energy feedback. Analyses aimed at
discussing the WHIM detectability from simulations, both in emission
\citep[e.g.,][]{yoshikawa04} and in absorption
\citep[e.g.,][]{Cen_etal01,viel05whim} have been so far based on
approximate descriptions of the pattern of chemical enrichment. We
will present in a future paper an observationally oriented analysis of
our simulations, aimed at quantifying how the WHIM properties can be
recovered under realistic observational conditions in the presence of
different feedback schemes. There is no doubt that the possibility of
performing high-resolution spectroscopy both in the X-ray (e.g., with
micro-calorimeters onboard of large collecting area X-ray telescopes)
and in the UV band (i.e., the now operating Cosmic Origin Spectrograph
onboard of Hubble Space Telescope) will provide a leap forward in the
study of the diffuse warm-hot baryons in the local universe. Cosmological
hydrodynamical simulations, like those presented in this paper, 
offer the natural interpretive framework for these future
observations, which will not only complete the census of baryons at
low redshift but also characterize their physical properties.

\section*{Acknowledgments.}
We would like to thank the anonymous referee for constructive comments
that helped improving the presentation of the results.  Numerical
computations have been performed at CINECA (``Centro
Interuniversitario del Nord Est per il Calcolo Elettronico'') and CPU
time has been assigned thanks to an INAF-CINECA grant (key and
standard projects), and through an agreement between CINECA and the
University of Trieste. We acknowledge useful discussions with
L. Zappacosta. This work has been partially supported by the INFN-PD51
grant, by the ASI-AAE and ASI-COFIS Theory Grants, and by the
PRIN-MIUR Grant ``The Cosmic Cycle of Baryons''.

\bibliography{master_whim.bib}

\end{document}